\documentclass[usenatbib]{mnras}
\usepackage{multirow}
\usepackage{multicol}
\usepackage{booktabs}
\usepackage{rotating}
\usepackage{colortbl}
\usepackage{color}
\usepackage[dvipsnames]{xcolor}
\usepackage[fleqn]{amsmath}
\usepackage{amssymb}
\usepackage{amsfonts}
\usepackage{verbatim}
\usepackage{scalefnt}
\usepackage[percent]{overpic}
\usepackage[T1]{fontenc} 
\usepackage{aecompl} 
\usepackage{ulem}
\usepackage{cleveref}

\newcommand\T{\rule{0pt}{2.8ex}}       
\newcommand\B{\rule[-1.4ex]{0pt}{0pt}} 

\title[Revised event rates for EMRIs and X-MRIs]{Revised event rates for extreme and extremely large mass-ratio inspirals}

\author[V\'{a}zquez-Aceves et al.]{Ver\'{o}nica V\'{a}zquez-Aceves,$^{1}$ Lorenz Zwick,$^{2}$ Elisa Bortolas,$^{3,\,4}$ Pedro~R. Capelo,$^{2}$\newauthor Pau Amaro Seoane,$^{5,\,6,\,7,\,1}$ Lucio Mayer$^{2}$ and Xian Chen$^{8,\,7}$\thanks{Corresponding author: xian.chen@pku.edu.cn}
\\
$^{1}$Institute of Applied Mathematics, Academy of Mathematics and Systems Science, Chinese Academy of Sciences, 100190 Beijing, China\\
$^{2}$Center for Theoretical Astrophysics and Cosmology, Institute for Computational Science, University of Zurich, Winterthurerstrasse\\ \ 190, CH-8057 Z\"urich, Switzerland\\
$^{3}$ Dipartimento di Fisica ``G. Occhialini'', Universit\'a degli Studi di Milano-Bicocca, Piazza della Scienza 3, I-20126 Milano, Italy \\
$^{4}$ INFN, Sezione di Milano-Bicocca, Piazza della Scienza 3, I-20126 Milano, Italy\\
$^{5}$Institute of Multidisciplinary Mathematics, Universitat Polit\`{e}cnica de Val\`{e}ncia, Spain\\
$^{6}$DESY Zeuthen, Germany\\ 
$^{7}$Kavli Institute for Astronomy and Astrophysics at Peking University, 100871 Beijing, China\\
$^{8}$Department of Astronomy, School of Physics, Peking University, 100871 Beijing, China}

\date{Accepted XXX. Received YYY; in original form ZZZ}

\pubyear{2021}

\begin{document}
\label{firstpage}
\pagerange{\pageref{firstpage}--\pageref{lastpage}}
\maketitle


\begin{abstract}
One of the main targets of the Laser Interferometer Space Antenna (LISA) is the detection of extreme mass-ratio inspirals (EMRIs) and extremely large mass-ratio inspirals (X-MRIs). Their orbits are expected to be highly eccentric and relativistic when entering the LISA band. Under these circumstances, the inspiral time-scale given by Peters' formula loses precision and the shift of the last-stable orbit (LSO) caused by the massive black hole spin could influence the event rates estimate. We re-derive EMRIs and X-MRIs event rates by implementing two different versions of a Kerr loss-cone angle that includes the shift in the LSO, and a corrected version of Peters' time-scale that accounts for eccentricity evolution, 1.5 post-Newtonian hereditary fluxes, and spin-orbit coupling. The main findings of our study are summarized as follows: (1) implementing a Kerr loss-cone changes the event rates by a factor ranging between 0.9 and 1.1; (2) the high-eccentricity limit of Peters' formula offers a reliable inspiral time-scale for EMRIs and X-MRIs, resulting in an event rate estimate that deviates by a factor of about 0.9 to 3 when compared to event rates computed with the corrected version of Peters' time-scale and the usual loss-cone definition.
(3) Event rates estimates for systems with a wide range of eccentricities should be revised. Peters' formula overestimates the inspiral rates of highly eccentric systems by a factor of about 8 to 30 compared to the corrected values. Besides, for e$_0 \lesssim$0.8, implementing the corrected version of Peters' formula would be necessary to obtain accurate estimates. 
\end{abstract}




\section{INTRODUCTION}\label{sec:Introduction}

Massive black holes (MBHs) at the centre of galaxies can capture compact objects -- such as stellar-mass BHs, white dwarfs (WDs), neutron stars (NSs), and even brown dwarfs (BDs) -- that can either suffer a direct plunge or slowly inspiral to the event horizon without being disrupted. The latter events are known as extreme mass-ratio inspirals (EMRIs) when the mass ratio $q$ between the compact object and the MBH is of the order of 10$^{-4}$, and as extremely large mass-ratio inspirals (X-MRIs) when $q \sim 10^{-8}$. During the inspiral process, the binary would emit low-frequency gravitational waves (GWs) that space-borne detectors like the Laser Interferometer Space Antenna (LISA) can detect \citep[][]{Barack_2004,Pau_2012eLISA,klein_2016, LISA2_2017,Pau_2018,Barack2019}, providing detailed information about the binary and the surrounding space-time that is impossible to obtain via electromagnetic observations. Such detections would allow testing general relativity with exceptional accuracy; therefore, it is crucial to accurately estimate the event rates and the characteristics of inspiral processes.

The eccentricity of an EMRI or X-MRI when entering the LISA band depends on the formation channel. There are two basic formation scenarios that result in two different eccentricity regimes: the two body relaxation driven decay and the so-called Hills mechanism. 

The first scenario involves a dynamical process in which two-body relaxation increases the eccentricity of an orbiting object such that, at the pericentre, the object passes so close to the MBH that the energy loss by GW emission becomes significant. Ideally, after a pericentre passage, the orbital parameters evolve exclusively by GW emission, resulting in a very eccentric EMRI or X-MRI in which the semimajor axis can be very large compared to the pericentre. However, the process is not that simple. Relativistic effects can become relevant at pericentre. Moreover, at the apocentre, the two-body relaxation process that initially brought the object into the desired orbit could either enlarge the pericentre, making the GW emission negligible, or deflect the object into the loss cone where the secondary object rapidly plunges into the MBH and is lost to the system after a single GW burst \citep[][]{Alexander_2003}. The loss cone is a region of phase space such that the angular momentum of the incoming object is not large enough to escape the MBH \citep[][]{Merritt2013}; it is defined by an angle $\widehat{\theta}_{\rm lc}$, known as the loss-cone angle, and depends on the position of the last stable orbit (LSO). For a successful inspiral, the compact object has to be ``immune'' to relaxation processes once it has reached a pericentre that is sufficiently close to the MBH to emit GWs, i.e. its merger time-scale ($T_{\rm GW}$) has to be shorter than or similar to the time needed by two-body relaxation to perturb its pericentre. This condition is fulfilled at a specific critical semimajor axis ($a_{\rm crit}$) that depends on several factors like the mass and spin of the MBH, the distribution of stars and compact objects around it, and the merger time-scale. The value of the critical semimajor axis is necessary to estimate event rates \citep[][]{Sigurdsson_1997,Hopman_2005,Pau_2007}.

The second scenario involves a binary system composed of at least one compact object orbiting at short distances from the MBH. If the gravitational force from the MBH acting on one of the binary components is larger than the binding energy, the binary is disrupted and one of the objects is captured by the MBH. If the captured object is a compact object, it could become an EMRI or X-MRI, as it can resist tidal disruption at such close distances. Its initial semimajor axis would be equal to the binary-disruption distance, which is smaller than the semimajor axis involved in the two-body relaxation formation process. As a result, the eccentricity of an EMRI or X-MRI formed by this process would be low when it enters the LISA band \citep[][]{Pau_2020}. This idea is based on the work of \citet{hypervel}, who predicted that the presence of an MBH in the Galactic centre would result in the disruption of binary systems composed of main-sequence stars: one of the stars would be tidally disrupted, while the other would be ejected at high velocities (up to $v > 4\times 10^3$~km~s$^{-1}$). These ``hyper-velocity stars'' were first discovered by \citet{Brown2005}, who detected a star leaving the Galaxy with velocity $\sim$700~km~s$^{-1}$, providing evidence of the existence of the process. However, the population of binary systems containing compact objects near an MBH is not well understood, impeding reliable event-rate estimations.

In this work, we obtain event rates ($\dot{\Gamma}_{\rm i}$) of EMRIs and X-MRIs formed by two-body relaxation around \citet{Schwarzschild_1916} and \citet{Kerr_1963} MBHs with a mass similar to that of Sgr~A$^{\ast}$ ($4.3 \times 10^{6 }$~M$_{\sun}$), by modifying two elements: the merger (or inspiral) time-scale and the loss-cone angle $\widehat{\theta}_{\rm lc}$.

Usually, the merger time-scale of a binary system is obtained with Peters' formula \citep[][]{Peters_Mathews_1963,Peters_1964}. However, in this formation scenario, the high eccentricities and the relativistic effects that appear in the proximity of the MBH reduce the accuracy of Peters' approach. In Section~\ref{sec:timescales}, we describe a set of correction factors presented by \citet{Zwick_2020,Zwick_2021} that improve Peters' time-scale behavior under such circumstances, and compare it with an alternative form of Peters' formula, valid for high eccentricities, previously used to obtain EMRI and X-MRI event rates \citep[][]{Hopman_2005,Pau_2013,Pau_2019}. 

The critical semimajor axis and the loss cone depend on the position of the LSO, which is constant for a non-spinning MBH and is defined through the Schwarzschild radius ($r_{\rm S} = 2GM/c^{2} $), but for a Kerr MBH it also depends on the spin magnitude and on the orbital inclination of the secondary object. \citet{Pau_2013} obtained event rates that account for this effect and found that objects originally classified as direct plunges can form an EMRI if they approach in prograde orbits, whereas objects in retrograde orbits contribute more to the plunge rate. In Section~\ref{sec:criticalsma}, we re-derive the critical semimajor axis for Schwarzschild and Kerr MBHs including the correction factors in the merger time-scale and the shift in the LSO. In Section~\ref{sec:evenrate}, we present the necessary elements to estimate the event rates, and we derive two versions of a Kerr loss-cone angle that account for the shift in the LSO position to finally obtain an expression for $\dot{\Gamma}_{\rm i}$ that includes the Peters' time-scale corrections and the Kerr loss-cone angle.

In Section~\ref{sec:effectinrates}, we analyze the effects of the time-scale correction factors and the Kerr loss-cone angle on the event rates for EMRIs and X-MRIs composed of stellar-mass BHs, NSs, WDs, and BDs. We consider prograde and retrograde orbits with orbital inclinations $|\theta|=[0, 0.1, 0.4, 0.7, 1.0, 1.3, 1.57]$ radians, and an MBH with dimensionless spin $a_{\bullet}=0$ to $a_{\bullet}=0.999$.\footnote{We define the MBH spin as $a_\bullet = c J_{\rm MBH} /(G M_{\rm MBH}^2)$, where $M_{\rm MBH}$ and $J_{\rm MBH}$ are the MBH's mass and angular momentum magnitude, respectively, $c$ is the speed of light in vacuum, and $G$ is the gravitational constant.} Finally, we conclude in Section~\ref{sec:conclussions}.


\section{THE INSPIRAL TIME-SCALE}\label{sec:timescales}

A reliable estimate of the merger time-scale is needed to understand the inspiral processes. In this work, we refer to a generic GW-induced decay time-scale as $T_{\rm{GW}}$. The most commonly used estimate of $T_{\rm{GW}}$ is the so-called Peters' formula,

\begin{align}
&T_{\rm P}(a_0,e_0)= \frac{5}{256} \frac{a_{\rm 0}^4 c^5}{G^3 M_{\rm MBH}\, m_2 (M_{\rm MBH}+m_2)} f(e_{\rm 0}), \label{eq:TP} \\
&f(e_{\rm 0})= (1-e_{\rm 0}^2)^{7/2} \left( 1 +\frac{73}{24} e_{\rm 0}^2 + \frac{37}{96} e_{\rm 0}^4 \right)^{-1}, \nonumber
\end{align}

\noindent where $a_0$ and $e_0$ are the initial semimajor axis and eccentricity, respectively, and $m_2$ is the mass of the orbiting compact object. This formula is obtained from the average change in the semimajor axis due to energy loss by GW emission, $\langle {\rm d}a/{\rm d}t \rangle$, described by \citet{Peters_Mathews_1963} based on the following assumptions:

\begin{itemize}
    \item[(A)] The binary's orbit is Keplerian.    
    \item[(B)] GW radiation is described by \citet{GR}'s quadrupole formula.    
    \item[(C)] The secular evolution of the orbital parameters is slow with respect to the period of the orbit.
\end{itemize}
Equation~\eqref{eq:TP} is obtained by integrating $\langle {\rm d}a/{\rm d}t \rangle$ assuming that
\begin{itemize}
    \item[(D)] The secular evolution of the eccentricity can be neglected.
\end{itemize}
Because of these assumptions, Peters' formula is not an exact measure of $T_{\rm{GW}}$, and fails to accurately model the behavior of highly eccentric and highly relativistic orbits. Note that \citet{Peters_1964} provides a merger time-scale valid for arbitrary eccentricities,
\begin{align}
T_{\rm GW}=& \frac{60}{1216} \frac{  c_0^4 \,c^5}{G^3 M_{\rm MBH}\, m_2 (M_{\rm MBH}+m_2)} \nonumber \\
&\times \int_{0}^{e_0} \frac{de \, e^{29/19} [1+(121/304)e^2]^{1181/2299}}{(1-e^2)^{3/2}},
\label{eq:TPallecc}
\end{align}
where c$_0$ is a constant obtained from the initial conditions a$_0$ and e$_0$. However, its solution requires numerical integration.

\begin{figure}
\vspace{-10.0pt}
\includegraphics[width=0.5\textwidth]{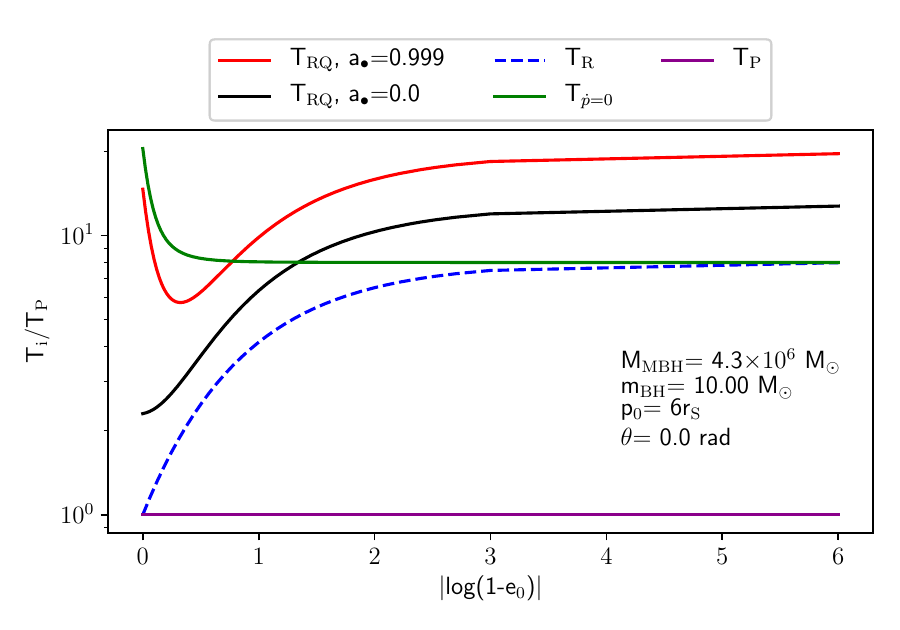}
\vspace{-15.0pt}
\caption{Comparison between the different merger time-scales normalized to the Peters' time-scale $T_{\rm P}$ (purple solid line), for the case of a compact object of 10~M$_{\sun}$ in an equatorial prograde orbit around a $4.3 \times 10^6$~M$_{\sun}$ MBH, and a pericentre distance $p_0 = 6 r_{\rm S}$. The green solid line, $T_{\dot{p}=0}$, is the time-scale obtained with Equation~\eqref{eq:TPapprox}. The blue dashed line, $T_{\rm R}$, includes the correction for the secular eccentricity evolution $R$ (Equation~\ref{eq:re}). The black ($a_\bullet = 0$) and red ($a_\bullet = 0.999$) solid lines, $T_{\rm RQ}$, include the correction factors $R$ and $Q$, which accounts for post-Newtonian effects up to order 1.5.}
\label{fig:Times}
\end{figure}

The case of highly eccentric orbits ($e_0>0.9$) is important for EMRI and X-MRI event rates estimates. Therefore, some authors \citep[e.g.][]{Hopman_2005,Pau_2013, Pau_2019} adopted a different estimate for $T_{\rm{GW}}$ which is obtained by modifying assumption (D) to
\begin{itemize}
    \item[ (D$^{\ast}$)] The secular evolution of the pericentre ($p_0$) can be neglected.    
\end{itemize}
The resulting time-scale follows from integrating Equation~\eqref{eq:TPallecc} over one period assuming $M_{\rm MBH} \gg m_2$, and $e\to1$ \citep{Peters_1964} 
\begin{align}
T_{\dot{p}=0}(a_0,e_0) \simeq \sqrt{2}\frac{24}{85} \frac{a_{\rm 0}^4 c^5}{G^3 m_2 M_{\rm MBH}^2} (1-e_0)^{7/2}.
\label{eq:TPapprox}
\end{align}
Assumption (D$^{\ast}$) is much better suited to describe orbits with extreme eccentricities for which the pericentre remains nearly constant during the inspiral. The time-scales given by Equations~\eqref{eq:TP} and \eqref{eq:TPapprox} differ in their range of validity because of their different assumptions.

In \citet{Zwick_2020}, a simple correction factor, $R$, is proposed to correct for the omitted secular eccentricity evolution of Equation \eqref{eq:TP}, and therefore it interpolates between the low-eccentricity validity of $T_{\rm P}$ and the high-eccentricity validity of $T_{\dot{p}=0}$ \citep[see also][for an alternative formulation]{Bonetti2018}. The correction factor is

\begin{align}
    R(e_0)= 8^{1-\sqrt{1-e_0}} .
\label{eq:re}
\end{align}

By multiplying $T_{\rm P}$ by $R$, Peters' formula extends its validity to all eccentricities. The corrected formula, denoted as $T_{\rm R}$, reproduces $T_{\dot{p}=0}$ in the high-eccentricity limit. The time-scale $T_{\rm R}$ effectively does away with assumption (D) or (D$^{\ast}$).

\citet{Zwick_2021} obtained an additional correction factor $Q$ that improves the estimate of the inspiral time-scale by modelling post-Newtonian effects up to order 1.5, based on the following assumptions:

\begin{itemize}

    \item[(A$^{\ast}$)] The binary's orbit is post-Newtonian (1~PN).
    
    \item[(B$^{\ast}$)] GW radiation is described by post-Newtonian fluxes (1.5 relative PN).
    
    \item[(C)\;\;] The secular evolution of the orbital parameters is slow with respect to the period of the orbit.
    
\end{itemize}

The common theme among all of these formulations is assumption (C), which is appropriate when the mass ratio of the binary is extreme and the residence time at a given separation, $a/\dot{a} \propto 1/q$, is much longer than the period of an orbit.

The explicit formula of $Q$ is found in \citet{Zwick_2021}; this correction factor is valid for arbitrary values of initial eccentricity, semimajor axis, MBH spin, and orbital inclination $\theta$. The fully corrected time-scale is

\begin{align}
 T_{\rm RQ} = T_{\rm P}R Q.
 \label{eq:T_pRQ}
\end{align}
\noindent

\noindent Equation~\eqref{eq:T_pRQ} takes into account the relativistic effects that can influence the inspiral resulting in an accurate time-scale for the eccentricity regime ($e_0>0.9$) and pericentre distances ($p_0\lesssim 6 r_{\rm S}$) expected for EMRIs and X-MRIs \citep[see also][where an improved inspiral time-scale for parabolic orbits based on numerical perturbative calculations is presented]{Gair_2006}.

\begin{table}
\centering
\begin{tabular}{c|c|c|c}
 \hline
 Name& Validity &Time-scale accuracy & PN effects \T \B \\
 \hline
 $T_{\rm P}$     & Low $e_0$  & underestimates & No \T \B \\
 $T_{\dot{p}=0}$ & High $e_0$ & over/under     & No \T \B \\
 $T_{\rm R}$    & Any $e_0$  & underestimates & No \T \B \\
 $T_{\rm RQ}$   & Any $e_0$  & best           & Yes \T \B \\
 \hline
\end{tabular}
\caption{In this table we summarize the characteristics and validity ranges of the different estimates for the GW-induced merger time-scales. We use the corrected time-scale $T_{\rm R Q}$ as a benchmark from which to say whether the time-scales $T_{\dot{p}=0}$, $T_{\rm P}$, and $T_{\rm R}$ over or underestimate the result.}
 \label{tab:GWtimescales}
\end{table}

Figure~\ref{fig:Times} shows the differences between the various GW-induced merger time-scales as a function of eccentricity for a BH of 10~M$_{\sun}$ in an equatorial prograde orbit with a pericentre distance, $p_0$, of 6 $r_{\rm S}$, around a Milky Way-like MBH, with M$_{\rm MBH}=4.3\times10^6 $~M$_{\sun}$ and $a_{\bullet}=0$ or $a_{\bullet}=0.999$. Note that, in the limit of high eccentricity, $T_{\rm P} \to T_{\dot{p}=0} /8$ and $T_{\dot{p}=0}$ is $\sim$60 ($\sim$40) per cent of $T_{\rm RQ}$ when $a_{\bullet}=0$ ($a_{\bullet}=0.999$). An interesting coincidence occurs when the correction factors $R$ and $Q$ multiplied together mimic the factor of $\sim$8 difference between $T_{\rm P}$ and $T_{\dot{p}=0}$ and the two lines ($T_{\rm RQ}$ and $T_{\dot{p}=0}$) cross over each other. For equatorial orbits, this effect occurs at $e_0\sim 0.8$ if $a_{\bullet}=0.999$ and at $e_0\sim$ 0.95 if $a_{\bullet}=0$.

Peters' time-scale $T_{\rm P}$ underestimates the merger time since it assumes that the eccentricity remains at its initial value throughout the evolution, artificially boosting the radiation of GWs. In contrast, the time-scale $T_{\dot{p}=0}$ overestimates the merger time-scale for low eccentricities since it assumes that the pericentre of the orbit does not decay, artificially decreasing the amount of GW emitted. The effect of the PN correction factors is to increase the estimate of the merger time-scale, especially for circular orbits. Nevertheless, for $p_0\lesssim 6 r_{\rm S}$ and eccentricity values $e_0 > 0.9$, relevant for EMRIs and X-MRIs, $T_{\dot{p}=0}\in (1.2 - 0.4)\times T_{\rm RQ}$. Table~\ref{tab:GWtimescales} summarizes the characteristics of the different merger time-scales.


\section{The critical semimajor axis}\label{sec:criticalsma}

The critical semimajor axis $a_{\rm crit}$ marks the end of the relaxation-driven evolution regime, i.e. two-body relaxation effects become irrelevant, and the energy loss due to GWs emission dominates the evolution of the orbit. Its value is found by solving

\begin{equation}
T_{\rm GW}=T_{\rm peri}(a_0, e_0),
\label{eq:TperiTinsp}
\end{equation}

\noindent where $T_{\rm GW}$ is the merger time-scale, and $T_{\rm peri}$ is the time required by two-body relaxation to change the pericentre of the inspiraling object \citep[][]{Pau_2007}, given by

\begin{align}
T_{\rm peri}(a_0, e_0)=T_{\rm rlx}(a_0) \times (1-e_0^2).
\label{eq:T_peri}
\end{align}

This time-scale is derived from the angular momentum diffusion time-scale,

\begin{align}
T_{\rm J}\sim T_{\rm rlx} \times [J/J_{\rm max}]^2 =T_{\rm rlx}(a) (1-e_0^2),
\label{eq:T_J}
\end{align}

\noindent where $J = (a(1-e^2)GM_{\rm MBH})^{1/2}$, $J_{\rm max} = (aGM_{\rm MBH})^{1/2}$ is the angular momentum of a circular orbit, and $T_{\rm rlx}(a)$ is the relaxation time-scale at a distance equal to the semimajor axis $a$ (see Equation~\ref{eq:Trlx} below).

Two-body interactions also cause energy diffusion, which changes the semimajor axis of the objects orbiting the MBH on the relaxation time-scale

\begin{align}
 T_{\rm rlx}\sim \mathcal{E}/\dot{ \mathcal{E}},
 \label{eq:T_E}
\end{align}

\noindent where $\mathcal{E}$=$GM_{\rm MBH}/2a$. This effect is not relevant to the inspiral process because, only by energy diffusion, it can take several relaxation time-scales to reach orbital parameters for which GWs emission becomes significant. From Equation~\eqref{eq:T_J}, if $e_0\neq0$, then $T_{\rm J}\lesssim T_{\rm rlx}$, indicating that two-body relaxation diffuses angular momentum faster than it diffuses energy \citep[][]{Hopman_2005}, increasing the eccentricity of an orbit while its semimajor axis remains approximately constant, allowing the object to reach very short distances to the MBH at the pericentre. This is a key dynamical aspect in the EMRIs/X-MRIs formation; taking $T_{\rm peri}(a_0, e_0)=T_{\rm J}$ in Equation~\eqref{eq:TperiTinsp} guarantees that objects with $a_0 \lesssim a_{\rm crit}$ that reached highly eccentric orbits due to a decrease in their angular momentum, merge before their pericentre changes by diffusion. If the term $(1-e_0^2)$ is omitted in the same equation, systems with longer $T_{\rm GW}$ for which diffusion in angular momentum is still significant can be mistakenly considered as potential inspiraling sources, artificially increasing the event rates.

At each pericentre passage, the energy loss by GWs emission is maximum, shrinking the semimajor axis and increasing the binding energy between the MBH and the compact object. If the new binding energy is high enough, the binary decouples dynamically from the surrounding stellar system, and diffusion in $J$ and $\mathcal{E}$ becomes negligible; the orbit then evolves only due to the energy loss by GWs emission, creating a successful inspiral that ends when the object crosses the event horizon of the central MBH. Otherwise, the object remains in the dynamical regime performing a random walk in the phase space driven by diffusion in energy and angular momentum, to either plunge into the MBH, diffuse to wider orbits, or become a potential inspiral source by diffusing into tighter orbits.\\ \\
For each given semimajor axis $a_0 \lesssim a_{\rm crit}$, there is a critical eccentricity above which $m_2$ becomes immune to the relaxation processes, and by fixing the pericentre of $m_2$ at the LSO around the MBH, we set a high-eccentricity limit given by $e_{\rm plunge}=1-r_{\rm LSO}/a_0$, where $r_{\rm LSO}$ is the position of the LSO. If for a given $a_0 \lesssim a_{\rm crit}$, $e_0>e_{\rm plunge}$, the pericentre of the orbit is located inside the LSO and $m_2$ crosses the event horizon of the MBH, without inspiraling, after a single pericentre passage \citep[][]{Pau_2007}. \\
This formation scenario is characterized by the balance between $T_{\rm peri}$ and the GW time-scale $T_{\rm GW}$. While the latter depends only on the source characteristics, the former requires a model of the stellar density in the vicinity of the MBH.\\ \\
We obtain $r_{\rm LSO}$ from the critical angular momentum described in \citet{ShapTeu_1983} for a non-relativistic particle in a highly eccentric orbit around a Schwarzschild BH, 
\begin{align}
J_{\rm crit}= 4GM_{\rm MBH}/c.
\label{eq:Jcrit}
\end{align}
The last equation describes a parabolic orbit with a pericentre distance equal to $p_{\rm LSO}=4r_{\rm S}$. As any particle with $J < J_{\rm crit}$ plunges into the MBH, the value of $p_{\rm LSO}$ defines the plunge radius. Therefore, for the Schwarzschild case, we assume that the LSO is located at the plunge radius $r_{\rm LSO}=4r_{\rm S}$.\\
For a Kerr MBH, the change in the position of the LSO due to the spin and inclination with respect to the spin axis, is modelled through the function $\mathcal{W} (\theta, a_{\bullet})$ derived in \citet{Pau_2013} from the separatrices between stable and unstable (plunging) orbits. The inclusion of $\mathcal{W} (\theta, a_{\bullet})$ can increase the number of cycles a prograde EMRI or X-MRI spends inside the LISA frequency band in such a way that it becomes detectable. Its derivation is based on the scheme presented in \citet{Sopuerta_2011}, that takes elements from the multipolar, post-Minkowskian formalism and BH perturbation theory to describe an inspiral trajectory, including also the radiation-reaction from the GWs emission to the system. For a Kerr MBH, the effective pericentre at the LSO is
\begin{equation}
p = a(1-e) = \mathcal{W} (\theta, a_{\bullet}) \times \frac{8 G M_{\rm MBH}}{c^2}.
\label{eq:periW}
\end{equation}

To estimate $T_{\rm rlx}$, we consider that inside the influence radius of the MBH, defined as $R_{\rm h}=GM_{\rm MBH}/\sigma_0^2$, with $\sigma_0$ the central velocity dispersion, the stellar density distribution follows a power-law cusp $\rho(r)\sim r^{-\gamma}$. This is a theoretical prediction of \citet{Peebles} and \citet{BW} from the 1970s that has been tested in the past decade by numerical approaches \citep[see, e.g.][]{Marc2001, Pau_2004, Preto_2009}, concluding that stellar cusps may be common around MBHs. A strong support of these assumptions is found in the work of \citet{Distribution1, Distribution2, Distribution3}, in which the authors conduct an extensive search for the stellar density cusp around Sgr~A$^{\ast}$, by performing observations and $N$-body simulations of the innermost structure of the Milky Way's nuclear star cluster. They find an excellent agreement between the theory, their observational data, and their simulations, consistent with the existence of a power-law density cusp around Sgr~A$^{\ast}$. 

Stellar-mass BHs dominate the central density as they sink to the centre due to mass segregation forming a cusp for which different indices have been suggested, for example $\gamma=1.3$--1.4 \citep[][]{Marc2006}, $\gamma=1.75$ \citep[][]{BW}, and $2\lesssim \gamma \lesssim 11/4$ in strong mass segregation scenarios \citep[][]{Alexander2009,Preto_2009,PauPreto2011}. We assume that these objects with a typical mass of $m_{\rm BH}=10$~M$_{\sun}$ are the driving species in the relaxation process. Less massive species distribute into a shallower profile and do not affect the relaxation rates. 

The relaxation time-scale inside $R_{\rm h}$ at a distance equal to the semimajor axis $a$ \citep{Baumgardt_2004a, Baumgardt_2004b, Marc_2002,Hopman_2005} is

\begin{align}
&T_{\rm rlx} = T_0\left(\frac{a}{R_{\rm h}}\right)^{\gamma-3/2},\label{eq:Trlx}  \\
&T_0=0.3389\, \frac{\sigma^3_0}{\text{ln}(\Lambda)  G^2 m_{\rm BH}^2 n_0 },\label{eq:T0} 
\end{align}

\noindent where ln$(\Lambda)\simeq 13$ is the Coulomb logarithm \citep[][]{BinneyTremaine_1987}, and $n_0$ is the number density given by

\begin{align}
n_0&=\frac{3-\gamma}{4\pi}\frac{N_0}{R^3_{\rm h}}, \\
\sigma_0 &=\left( \frac{1}{1+\gamma}\frac{G M_{\rm MBH}}{R_{\rm h}} \right)^{1/2},
\end{align}

\noindent where $N_0$ is the number of stellar-mass BHs inside $R_{\rm h}$. With these elements, Equation~\eqref{eq:T0} becomes

\begin{align}
T_0 \simeq & \frac{4.26}{(3-\gamma)(1+\gamma)^{3/2}} \frac{\sqrt{R_{\rm h}^{3}(GM_{\rm MBH})^{-1}}}{\text{ln}(\Lambda) N_0}\left( \frac{M_{\rm MBH}}{m_{\rm BH}} \right)^2. \label{eq:T0_complete}
\end{align}
 
\begin{figure}
\vspace{-12.0pt}
\includegraphics[width=0.5\textwidth]{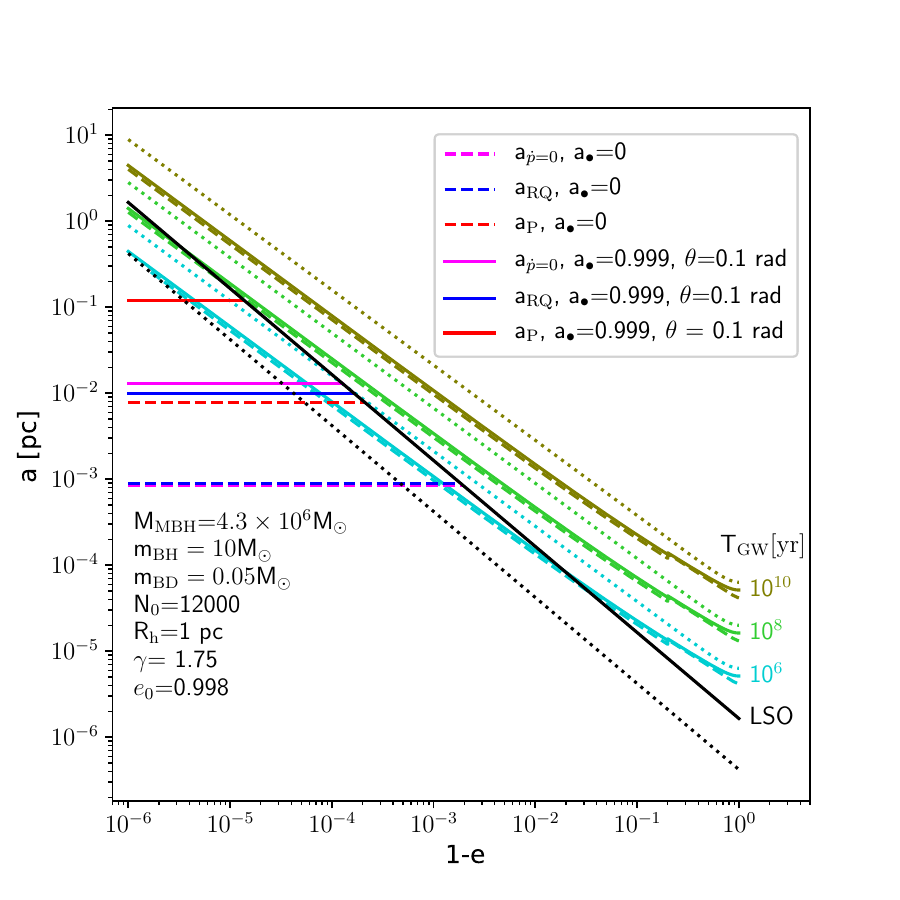}
\vspace{-15.0pt}
\caption{Critical semimajor axis as a function of eccentricity, for an inspiraling BD with $e_{0}$=0.998. We consider $a_{\bullet}$=0 and $a_{\bullet}$=0.999, for which the orbital inclination of the BD is $\theta$=0.1 rad. These values of spin and $\theta$ maximize the correction associated with the spin.  The diagonal green colored curves are isochrones that represent the inspiral time of the binary in years. The dotted isochrones are obtained with Peters' formula (Equation~\ref{eq:TP}), the solid isochrones with the corrected time-scale $T_{\rm RQ}$ and $a_{\bullet}$=0, and the dashed isochrones with $T_{\rm RQ}$ and $a_{\bullet}$=0.999. Black lines represent the LSO, the solid line is the Schwarzschild case, and the dotted, the Kerr case. The intersection between the LSO and the horizontal lines are the values of $a_{\rm crit}$ for the Schwarzschild and Kerr cases. The subscript $RQ$ (blue lines) indicates that the correction factors in Equation~\eqref{eq:acrit} are included, whereas the subscript $P$ (red lines) indicates the value without the corrections. The magenta lines indicates the value of $a_{\dot{p}=0}$, obtained using the merger time-scale $T_{\dot{p}=0}$, given by Equation~\eqref{eq:TPapprox}.}
\label{fig:acrit}
\end{figure}

As inspiral time-scale, we take Equation~\eqref{eq:T_pRQ} to include the correction factors, and evaluate it at the LSO as long as it is located no closer than $3r_{\rm S}$. At shorter distances, the correction factors involving the PN terms are not accurate \citep[][]{Zwick_2021}; in that case, we compute $Q$ at $3r_{\rm S}$. We re-write the correction factor $Q$ in terms of the MBH spin, the function $\mathcal{W} (\theta,a_{\bullet})$, and the initial eccentricity.

If $ p_{\rm 0} \leq 3 r_{\rm S}$, the correction factor Q is computed at $3r_{\rm S}$:

\begin{align}
Q= &\exp(A) \left[1.67-2.75\left(e_0^2-3.16e_0\right)\right],  \label{eq:QALSO} \\
A=& s\left[\frac{0.15}{1.5} e_0 + 1.37(1-e_0)^{3/2}\right] + \nonumber \\ 
&+|s|^{3/2} \left[\left(0.36 e_0\right)^{5/2} + 2.94(1-e_0)^3\right],\nonumber
\end{align}

\noindent where $s = a_{\bullet}\cos(\theta)$.

If $ p_{\rm 0} > 3 r_{\rm S}$, the correction factor Q is given by 

\begin{equation}
Q=q_{\rm h}\left(e_0,\mathcal{W}(\theta,a_{\bullet})\right)q_{\rm s}\left(s,e_0,\mathcal{W} (\theta, a_{\bullet})\right),
\label{eq:Qe}
\end{equation}

\noindent with

\begin{align}
&q_{\rm h}(e_0,\mathcal{W}(\theta,a_{\bullet}))= \nonumber \\ 
&\exp\Big(\frac{0.7}{\mathcal{W}(\theta,a_{\bullet})}\Big) \times\Bigg\{1+\Big(1-e_0\Big)^2\left[\exp\left(\frac{0.55}{\mathcal{W}(\theta,a_{\bullet})}\right)-1\Big)\right] +\nonumber \\ 
&+\left(\frac{0.95}{\mathcal{W}(\theta,a_{\bullet})}\right)^{3/2}\left(e_0^2-e_0\right)\Bigg\},\nonumber
\end{align}

\begin{align}
&q_{\rm s}\left(s,e_0,\mathcal{W}(\theta,a_{\bullet})\right)= \nonumber \\ 
&\exp \Bigg\{ s \Bigg[ \frac{0.3 \, e_0}{4\mathcal{W} (\theta, a_{\bullet})} + \left(1-e_0\right)^{3/2} \left( \frac{3.7}{4\mathcal{W} (\theta, a_{\bullet})} \right)^{3/2}\Bigg]+ \nonumber \\ 
&+ |s|^{3/2} \Bigg[\left(\frac{0.275 \, e_0}{\mathcal{W} (\theta, a_{\bullet})} \right)^{5/2} +(1-e_0)^3\left(\frac{1.075}{\mathcal{W} (\theta, a_{\bullet})} \right)^3\Bigg] \Bigg\}.\nonumber
\end{align}

Combining these elements, the critical semimajor axis takes the form

\begin{align}
a_{\rm crit}=& R_{\rm h} \left[\varepsilon\; \mathcal{W}(\theta, a_{\bullet})^{5/2} \frac{m_{\rm BH}^2}{m_2(M_{\rm MBH} + m_2)}\right] ^{1/(\gamma - 3)} \times \nonumber \\
&\times \left[\frac{f(e_0)}{(1-e_0^2)(1-e_0)^{5/2}}\; R\,Q\right]^{1/(\gamma - 3)}, \label{eq:acrit}\\ 
\varepsilon=&\frac{(3-\gamma)(1+\gamma)^{3/2}}{4.26} \left(\frac{5\times 8^{5/2}}{256}\right)N_0 \text{ln}(\Lambda). \nonumber
\end{align}

The correction factors can be turned off by setting $RQ=1$, in which case $a_{\rm crit}$ is determined by $T_{\rm P}$ (Equation~\ref{eq:TP}) and is denoted as $a_{\rm P}$. If the correction factors $RQ$ are applied, the critical semimajor axis is denoted as $a_{\rm RQ}$, and as $a_{\dot{p}=0}$ if its value is derived using $T_{\dot{p}=0}$ as the merger time-scale (see equation~28 in \citealt{Pau_2019}). 

Figure~\ref{fig:acrit} shows the values of $a_{\rm RQ}$, $a_{\rm P}$, and $a_{\dot{p}=0}$ for a BD X-MRI with $e_0$=0.998 inspiraling into a Schwarzschild MBH, and into a Kerr MBH for which $a_{\bullet}$=0.999, and $\theta$=0.1 rad. For $a_{\bullet}$=0, we find that $a_{\rm P} \sim 8.5~a_{\rm RQ}$, which can significantly reduce the event rates when computed with $a_{\rm RQ}$; in contrast, the value of $a_{\dot{p}=0}$ is of about 0.9 $a_{\rm RQ}$. In the Kerr case, the effect of $R$ and $Q$ is larger resulting in  $a_{\rm P} \sim 12~a_{\rm RQ}$, and $a_{\dot{p}=0}\sim 1.3 ~a_{\rm RQ}$. 

The difference between $a_{\rm P}$ and the values obtained for $a_{\dot{p}=0}$ and $a_{\rm RQ}$ originates in the lack of accuracy of Peters' time-scale for highly eccentric and relativistic orbits, revealing the need to thoroughly verify Peters' time-scale's validity depending on the physical characteristics of the system.


\section{THE INSPIRAL EVENT RATE}\label{sec:evenrate}

The event rate of successful inspirals is calculated by integrating the number of sources \citep[][]{Hopman_2005}, $n(a)$, in a volume defined from the minimum distance at which we expect to find at least one potential EMRI/X-MRI source, $a_{\rm min}$ \citep[][see Equation~\ref{eq:a_min} below]{Pau_2019}, to the critical semimajor axis:

\begin{equation}
\dot{\Gamma}_{\rm i} \simeq \int_{a_{\rm min}}^{a_{\rm crit}} \frac{{\rm d}n(a)}{T_{\rm rlx} (a)\; \text{ln}(\widehat{\theta}_{\rm lc}^{-2})},
\label{eq:Rates}
\end{equation}

\noindent where $dn(a)$ is obtained from the number of potential sources around the MBH, as explained below, and $\widehat{\theta}_{\rm lc}$ is the loss-cone angle associated to the position of the LSO. 

Inside the integration volume defined by $a_{\rm min}$ and $a_{\rm crit}$, two-body relaxation is the leading mechanism that brings a source sufficiently close to the MBH to produce an inspiral event. In the following section, we derive $a_{\rm min}$, $dn(a)$, and the loss-cone angle of a Kerr MBH that can deviate from a Schwarzschild loss-cone angle if the spin of the MBH is sufficiently high. 

\subsection{Number of sources around a MBH}

\begin{table}
\centering
\renewcommand{\arraystretch}{1.2}
 \begin{tabular}{c|c|c}
\hline 
Inspiraling object & mass [M$_{\sun}$] & $f_{\rm sub}$ \T \B \\
\hline
Stellar-mass BH & 10.0 & $8.13 \times 10^{-4}$ \T \B \\
Neutron star (NS)       & 2.7  & $4.24 \times 10^{-3}$ \T \B \\
White dwarf (WD)        & 0.8  & $7.20 \times 10^{-2}$ \T \B \\
Brown dwarf (BD)        & 0.05 & 0.21 \T \B \\ 
\hline 
\end{tabular} 
\caption{Mass and fraction number of inspiraling objects.}
\label{tab:EX-MRI_masses}
\end{table}

To obtain the number of objects of a given species, we assume that, similarly to stellar-mass BHs, lighter objects follow a mass density distribution given by a power law with exponent $\beta$, resulting in a two-population system in which the value of $\gamma$ dictates the distribution of the stellar-mass BHs around the MBH, and $\beta$ the inspiraling object's population distribution. The number of objects of a given species within a given semimajor axis is

\begin{align}
N(a)=f_{\rm sub} N_{\rm tot} \left( \frac{a}{R_{\rm h}}\right)^{3-\beta},
\label{eq:N(a)}
\end{align}

\noindent where $N_{\rm tot}$ is the total number of objects (main-sequence stars, compact objects, and substellar objects) within the influence radius of the MBH, and $f_{\rm sub}$ is the fraction number of the considered species obtained from a \citet{Kroupa2001} broken power law, $\Phi(m)\propto m_{\ast}^{-\alpha}$, with $m_{\star}$ the average stellar mass. We use $\alpha=[0.3, 1.3, 2.3]$ for the mass intervals $[0.01, 0.07, 0.5, 150] \times$~M$_{\sun}$ where [0.01--0.07]~M$_{\sun}$ is the BD mass range. Table~\ref{tab:EX-MRI_masses} shows the masses and fraction numbers of each considered object. The numerator of Equation~\eqref{eq:Rates} comes from differentiating Equation~\eqref{eq:N(a)}

\begin{align}
{\rm d}n(a)= f_{\rm sub}(3-\beta) \frac{N_{\rm tot}}{R_{\rm h}} \left( \frac{a}{R_{\rm h}}\right)^{2-\beta} da.
\label{eq:dna}
\end{align}

\noindent The distance $a_{\rm min}$ at which at least one object of a given species can be found is obtained by setting Equation \eqref{eq:N(a)} equal to 1 and taking $N_{\rm tot}=M_{\rm MBH}/m_{\star}$,

\begin{align}
a_{\rm min}=R_{\rm h}\left( \frac{m_{\star}}{f_{\rm sub} \, M_{\rm MBH}} \right)^{1/(3-\beta)}.
\label{eq:a_min}
\end{align}

\subsection{The loss-cone angle for Schwarzschild and Kerr black holes}\label{sec:LossConeAngle}

Not all the objects that approach the MBH can produce an EMRI or X-MRI. The pericentre distance has to be sufficiently small for GWs emission to occur; however, if it falls within the LSO, the object suffers a direct plunge. We can identify plunging orbits through their velocity vector; if it lies within a cone defined by a half-angle equal to the loss-cone angle, the orbit takes the object inside a sphere of radius equal to the LSO and rapidly merges with the MBH. The loss-cone angle is calculated as 

\begin{equation}
\widehat{\theta}_{\rm lc}=\left(\frac{J_{\rm max}}{J_{\rm lc}}\right)^{-1/2},
\label{eq:LossCone}
\end{equation}

\noindent where $J_{\rm lc}^2 $ is the angular momentum of an orbit around the MBH that takes a particle to a distance equal to the loss-cone radius $r_{\rm lc}$. For a Schwarzschild BH, we take $J_{\rm lc}= J_{\rm crit}$ from Equation \eqref{eq:Jcrit}, so that $r_{\rm lc,S} = 4 r_{\rm S}$, thus the associated loss-cone angle is
\begin{align}
\widehat{\theta}_{\rm S}=\left(\frac{a}{8 r_{\rm S}}\right)^{-1/4}.
\label{eq:LossConeS}
\end{align}

For a Kerr BH, the shift in the LSO changes the loss-cone radius; therefore, it is necessary to implement a Kerr loss-cone angle that considers this effect. 
We implement two versions for the Kerr loss-cone angle; the first, is based on the LSO position shift given by the function $\mathcal{W}$ ($\theta$, $a_{\bullet}$). We write the Kerr loss-cone radius as $r_{\rm lc,\mathcal{W}} = 4 r_{\rm S} \mathcal{W}$ ($\theta$, $a_{\bullet}$); the associated loss-cone angle is

\begin{align}
\widehat{\theta}_{\mathcal{W}}=\left(\frac{a}{8 r_{\rm S} \mathcal{W}(\theta, a_{\bullet})}\right)^{-1/4}= \widehat{\theta}_{\rm S}\mathcal{W}(\theta, a_{\bullet})^{1/4}
\label{eq:LossConeW}
\end{align}

\begin{figure}
\includegraphics[width=0.5\textwidth]{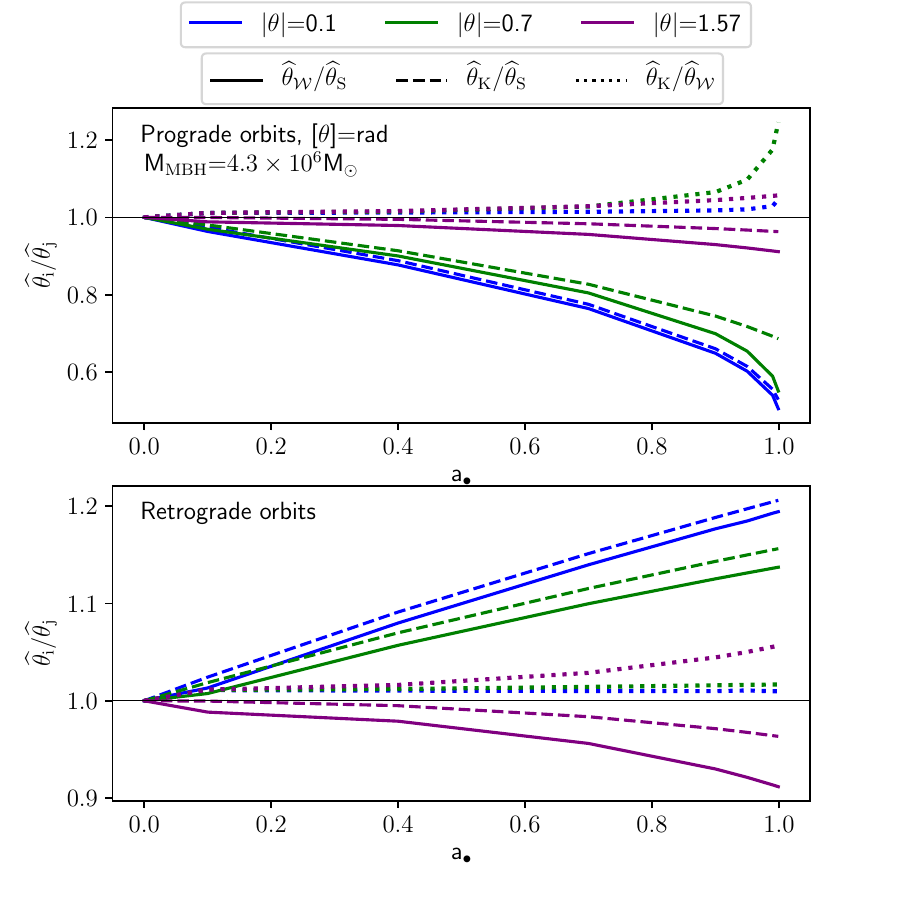}
\vspace{-15.0pt}
\caption{The figure shows the ratio between the loss-cone angles $\widehat{\theta}_{\rm i}/ \widehat{\theta}_{\rm j}=[\widehat{\theta}_{\rm \mathcal{W}}/\widehat{\theta}_{\rm S}, ~\widehat{\theta}_{\rm K}/ \widehat{\theta}_{\rm S},~\widehat{\theta}_{\rm K}/\widehat{\theta}_{\rm \mathcal{W}}]$ obtained with Equations~\eqref{eq:LossConeS}, \eqref{eq:LossConeW}, and \eqref{eq:LossConeK}. We show prograde (top panel) and retrograde (bottom panel) orbits with three different orbital inclinations: $|\theta|=[0.1, 0.7, 1.57]$~rad. The black horizontal line is plotted as a reference: the closer the lines are to the black line, the closer the values are between them. The ratios $\widehat{\theta}_{\rm \mathcal{W}}/\widehat{\theta}_{\rm S}$ and $\widehat{\theta}_{\rm K}/ \widehat{\theta}_{\rm S}$ indicate the deviation of the Kerr loss-cone angle with respect to the Schwarzschild case as a function of the MBH spin, whereas $\widehat{\theta}_{\rm K}/\widehat{\theta}_{\rm \mathcal{W}}$ shows the difference between the two versions of the Kerr loss-cone angle.}
\label{fig:LossConeWk}
\end{figure}

For the second version, we use the analytic approximation for the critical angular momentum ($J_{\rm crit}$) of a non-relativistic test particle orbiting a Kerr BH obtained by \citet{will2012}, which compared to numerical solutions is accurate within a 5 per cent for $0.9 \leq a_{\bullet} \leq 0.99$, and is valid for arbitrary orbital inclinations:

\begin{align}
J_{\rm crit} = &\frac{GM}{c}\left[2 K(\theta,a_{\bullet})+ 2 \right],
\label{eq:LcritkKerr} \\ \nonumber \\
K(\theta,a_{\bullet})=& \sqrt{1-s-(1/8)a_{\bullet}^2\sin^2(\theta) F(a_{\bullet},\cos(\theta)}), \nonumber \\
F(a_{\bullet},\cos(\theta)) =& 1+\frac{s}{2}+\frac{a_{\bullet}^2}{64}\left(7 + 13 \cos^2(\theta)\right)+ \nonumber \\ &+\frac{a_{\bullet}^2s}{128} \left(23 + 5 \cos^2(\theta)\right)+ \frac{a_{\bullet}^4}{2048} \Big(55 + \nonumber \\ &+ 340 \cos^2(\theta) - 59 \cos^4(\theta)\Big) + \mathcal{O}^5(a_{\bullet}^5). \nonumber
\end{align}

A particle with $J<J_{\rm crit}$ rapidly plunges into the MBH, determining a Kerr loss-cone angle given by
\begin{align}
\widehat{\theta}_{\rm K}=\left(\frac{2 a}{r_{\rm S}}\left[2K(\theta,a_{\bullet}) +2\right]^{-2}\right)^{-1/4}.
\label{eq:LossConeK}
\end{align}

An inspiraling object approaching an MBH in a prograde orbit finds the LSO closer to the MBH; consequently, the loss-cone angle magnitude decreases with respect to the Schwarzschild case, and it is easier for the incoming body to avoid direct plunge. For retrograde orbits, the LSO is pushed away from the MBH; hence the loss-cone angle magnitude is larger than in the prograde cases, increasing the phase space that produces a direct plunge. This effect can be easily seen in Equation~\eqref{eq:LossConeW} as $\mathcal{W} (\theta, a_{\bullet}) \lesssim 1$ for prograde orbits, and $\gtrsim 1$ for the retrograde cases. 

In Figure~\ref{fig:LossConeWk}, we show that the two versions of the Kerr loss-cone angle yield similar results, especially for retrograde orbits with $\theta\lesssim -1.3$~rad. The value of $\widehat{\theta}_{\rm K}$ slightly deviates from $\widehat{\theta}_{\rm \mathcal{W}}$ in the case of prograde orbits with $\theta \sim 0.4$--0.7 rad and $a_{\bullet} \gtrsim 0.95$.  

The MBH spin effects are weak for $\theta\sim\pm 1.57$~rad, and both versions of the Kerr loss-cone angle indicates that if the inspiraling object approaches an MBH in a highly inclined orbit, regardless if it is in a prograde or retrograde orbit, the phase space that produces a direct plunge is reduced if the MBH is rotating. The strongest spin effects appear with $\theta = \pm 0.1$~rad and $a_{\bullet}=0.999$; in the case of prograde orbits, $\widehat{\theta}_{\rm \mathcal{W}} \sim \widehat{\theta}_{\rm K} \sim 0.5 \,\widehat{\theta}_{\rm S}$, and $\widehat{\theta}_{\rm \mathcal{W}} \sim \widehat{\theta}_{\rm K} \sim 1.19 \,\widehat{\theta}_{\rm S}$ for $\theta=-0.1$~rad.

Finally, we use Equations~\eqref{eq:Trlx}, \eqref{eq:dna}, and \eqref{eq:LossCone} to solve Equation~\eqref{eq:Rates} with the correction factors embedded in the value of a$_{\rm crit}$, thus obtaining

\begin{multline}
\dot{\Gamma}_{\rm i} = \frac{3-\beta}{2\lambda T_0} \;\frac{ N_{tot}}{R_h^{\lambda}}\; f_{\rm sub} \; \times\\ \\
\times \left\lbrace 
\left[ a_{\rm crit}^{\lambda}\left(\text{ln}\left(\frac{a_{\rm crit}}{D}\right)- \frac{1}{\lambda}\right)\right]- \left[ a_{\rm min}^{\lambda} \left(\text{ln}\left(\frac{a_{\rm min}}{D}\right)- \frac{1}{\lambda} \right) \right] \right\rbrace,
\label{eq:EventRate}
\end{multline}

\noindent with $\lambda=(9/2)-\beta - \gamma$, and D a term associated to the loss-cone angle, given by

\begin{equation}
D=
\begin{cases}
8 r_{\rm S}, & \text{for $\widehat{\theta}_{\rm lc}^{-2}=\widehat{\theta}_{\rm S}^{-2}$,}\\ \\
8 r_{\rm S} \mathcal{W}(\theta,\rm a_{\bullet}), &\text{for $\widehat{\theta}_{\rm lc}^{-2}=\widehat{\theta}_{\mathcal{W}}^{-2}$,}\\ \\
0.5 r_{\rm S} \left[2 + 2 K(\theta,a_{\bullet})\right]^2,& \text{for $\widehat{\theta}_{\rm lc}^{-2}=\theta^{-2}_{\rm K}$.}
\end{cases}
\label{eq:D}
\end{equation}


\section{EFFECT OF THE CORRECTION FACTORS AND THE LOSS-CONE ANGLE IN THE EVENT RATES} \label{sec:effectinrates}

\begin{table}
\centering
\begin{tabular}{c|c|c|c|c}
\hline
\multicolumn{5}{c}{$a_{\bullet}=0.0$} \T \B \\
\hline 
Object&\multicolumn{2}{c|}{  $e_{\rm min}$ } & \multicolumn{2}{c}{ $e_{\rm max}$} \T \B \\ 
\hline
BD &    \multicolumn{2}{c|}{ 0.963178} & \multicolumn{2}{c}{ 0.998160} \T \B \\ 
\hline 
WD & \multicolumn{2}{c|}{ 0.981957} &\multicolumn{2}{c}{ 0.999790} \T \B \\ 
\hline 
NS & \multicolumn{2}{c|}{ 0.997268}&\multicolumn{2}{c}{ 0.999919} \T \B \\ 
\hline 
BH & \multicolumn{2}{c|}{ 0.996791} &\multicolumn{2}{c}{  0.999971} \T \B \\ 
\hline  \hline
\multicolumn{5}{c}{$a_{\bullet}=0.999$} \T \B \\
\hline 
& \multicolumn{2}{c|}{Prograde } & \multicolumn{2}{c}{Retrograde} \T \B \\
\hline
& $e_{\rm min}$ & $e_{\rm max}$ & $e_{\rm min}$ & $e_{\rm max}$ \T \B \\ 
\hline
BD & 0.990605 & 0.999954 &  0.947489 & 0.995773 \T \B \\ 
\hline 
WD & 0.995396 & 0.999995 & 0.974270 & 0.999505 \T \B \\ 
\hline 
NS & 0.999303 & 0.999998 & 0.996104 & 0.999810 \T \B \\ 
\hline 
BH & 0.999181 & 0.999999 & 0.995424 & 0.999932 \T \B \\
\hline
\end{tabular} 
\caption{Maximum and minimum eccentricity (Equations~\ref{eq:e_max} and \ref{eq:e_min}) for each type of compact object. The values of $e_{\rm min}$ and $e_{\rm max}$ are obtained for $\theta =  0.1$~rad (prograde orbits), $\theta =  -0.1$~rad (retrograde orbits ), $a_{\rm crit} = a_{\rm RQ}$, M$_{\rm MBH}= 4.3 \times 10^6$~M$_{\sun}$, and $a_{\bullet}=0$ or 0.999.}
\label{tab:ecc_co}
\end{table}
Applying the correction factors results in longer merger time-scales compared to $T_{\rm P}$ and $T_{\dot{p}=0}$ (for high enough eccentricities), giving more time to the relaxation processes to perturb the orbit of the compact object and prevent an inspiral. Therefore, to decouple from the dynamics regime, the inspiraling body has to be closer to the MBH, resulting in a smaller critical semimajor axis such that $a_{\rm RQ} \lesssim a_{\dot{p}=0}<a_{\rm P}$. The semimajor axis $a_0$ of an inspiraling object would be delimited by $a_{\rm min}$ and $a_{\rm crit} < R_{\rm h}$, and the pericentre distance would be fixed at the LSO position (Equation~\ref{eq:periW}). With these conditions, we define an eccentricity range $e_0=[e_{\rm min}(a_{\bullet},{\theta}),e_{\rm max}(a_{\bullet},{\theta})]$ given by

\begin{align}
&e_{\rm max}(a_{\bullet},{\theta})= 1-r_{\rm LSO}/a_{\rm crit}, \label{eq:e_max}\\
&e_{\rm min}(a_{\bullet},{\theta})= 1-r_{\rm LSO}/a_{\rm min}. \label{eq:e_min}
\end{align}

Objects with $e_0>e_{\rm max}$ plunge into the MBH after a single pericentre passage, and as we do not expect to find objects with $a_0<a_{\rm min}$, the value of $e_{\rm min}$ defines the plunging limit for the objects that are located closest to the MBH.

As the pericentre is fixed, orbits with $a_0=a_{\dot{p}=0}$, and specially with $a_0=a_{\rm P}$ can have higher eccentricities than orbits with $a_{0}=a_{\rm RQ}$. For this reason, we obtain $e_{\rm max}$ by setting $a_{\rm crit}=a_{\rm RQ}$ in Equation~\eqref{eq:e_max}; this eccentricity value is valid for all the cases, as $a_{\rm crit}$ sets the upper limit for the semimajor axis of an inspiraling orbit. Table~\ref{tab:ecc_co} shows the range of eccentricities for the considered compact objects obtained for $\theta = \pm 0.1$~rad, $a_{\bullet}$=[0.0, 0.999], and $a_{\rm crit}=a_{\rm RQ}$.

For the mass density distribution, we choose $\gamma=1.75$ for the stellar-mass BHs, $\beta=1.5$ for the lighter populations \citep{BW}, $R_{\rm h}=1$~pc, and $N_0=1.2\times 10^4$. We focus on systems composed of a central MBH of $4.3 \times 10^6$~M$_{\sun}$ with a spin range $a_{\bullet}=[0.0$--0.999] and an inspiraling object (a stellar-mass BH, an NS or a WD in the case of EMRIs; a BD in the case of X-MRIs) of mass $m_2$; the masses and fraction numbers are given in Table~\ref{tab:EX-MRI_masses}. Although we focus on MBHs with $M=4.3 \times 10^6$~M$_{\sun}$, we include event rate estimates for BH EMRIs with a central MBH of mass ranging from $10^4$ to $10^7$~M$_{\sun}$, which can be relevant for LISA. 

\begin{table}
\centering
\begin{tabular}{c||c|c||c|c}
\hline
\addlinespace[0.05cm]
$a_{\bullet}=0.0$& \multicolumn{2}{c||}{$\dot{\Gamma}_{\rm P} / \dot{\Gamma}_{\rm RQ}$} &\multicolumn{2}{c}{$\dot{\Gamma}_{\dot{p}=0}/ \dot{\Gamma}_{\rm RQ}$} \T \B \\
\hline
 &$e_{\rm min}$&$e_{\rm max}$ &$e_{\rm min}$&$e_{\rm max}$ \T \B \\
\hline  \hline
BD& 14.429 & 21.712 &1.364 & 2.057 \T \B \\ 
\hline 
WD& 15.319 & 20.628 &1.566 & 2.110 \T \B \\ 
\hline 
NS&  18.360 & 20.408 &1.913 & 2.127 \T \B \\ 
\hline 
BH &10.436 & 11.500 &1.675 & 1.846 \T \B \\ 
\hline
\addlinespace[0.6cm]
\end{tabular} 

\begin{tabular}{c||c|c||c|c}
\hline 
\addlinespace[0.05cm]
$a_{\bullet}=0.999$ & \multicolumn{4}{c}{$\dot{\Gamma}_{\rm P} / \dot{\Gamma}_{\rm RQ}$} \T \B \\
\hline
& \multicolumn{2}{c||}{$\theta=0.1$ rad}& \multicolumn{2}{c}{$\theta=-0.1$ rad} \T \B \\ 
\hline
 & $e_{\rm min}$&$e_{\rm max}$ &$e_{\rm min}$&$e_{\rm max}$ \T \B \\
\hline  \hline
BD & 26.021 & 32.344 & 10.365 & 16.295 \T \B \\ \hline
WD & 26.396 & 30.533 & 11.234 & 15.751 \T \B \\ \hline
NS & 28.609 & 30.205 & 14.031 & 15.794 \T \B \\ \hline
BH & 14.990 & 15.788 & 8.366 & 9.396 \T \B \\ \hline
\addlinespace[0.3cm]
\hline
\addlinespace[0.05cm]
$a_{\bullet}=0.999$ &\multicolumn{4}{c}{$\dot{\Gamma}_{\dot{p}=0}/ \dot{\Gamma}_{\rm RQ}$} \T \B \\
\hline 
 & \multicolumn{2}{c||}{$\theta=0.1$ rad}& \multicolumn{2}{c}{$\theta=-0.1$ rad} \T \B \\ 
\hline
 & $e_{\rm min}$&$e_{\rm max}$ &$e_{\rm min}$&$e_{\rm max}$ \T \B \\
\hline  \hline
BD & 2.693 & 3.349 & 0.936 & 1.477 \T \B \\ \hline
WD & 2.838 & 3.283 & 1.125 & 1.579 \T \B \\ \hline
NS & 3.114 & 3.288 & 1.433 & 1.614 \T \B \\ \hline
BH & 2.496 & 2.629 & 1.323 & 1.486 \T \B \\ \hline
\end{tabular} 
\caption{Ratios $\dot{\Gamma}_{\rm P}/ \dot{\Gamma}_{\rm RQ}$ and $ \dot{\Gamma}_{\dot{p}=0}/ \dot{\Gamma}_{\rm RQ}$ for EMRIs and X-MRIs with $\theta = \pm 0.1$~rad, around an MBH of mass M$_{\rm MBH}= 4.3 \times 10^6$~M$_{\sun}$, and $a_{\bullet}=0$ or 0.999.}
\label{tab:ratioRQ0}
\end{table}
\subsection{The effect of the correction factors} 

\begin{figure}
\includegraphics[width=0.5\textwidth]{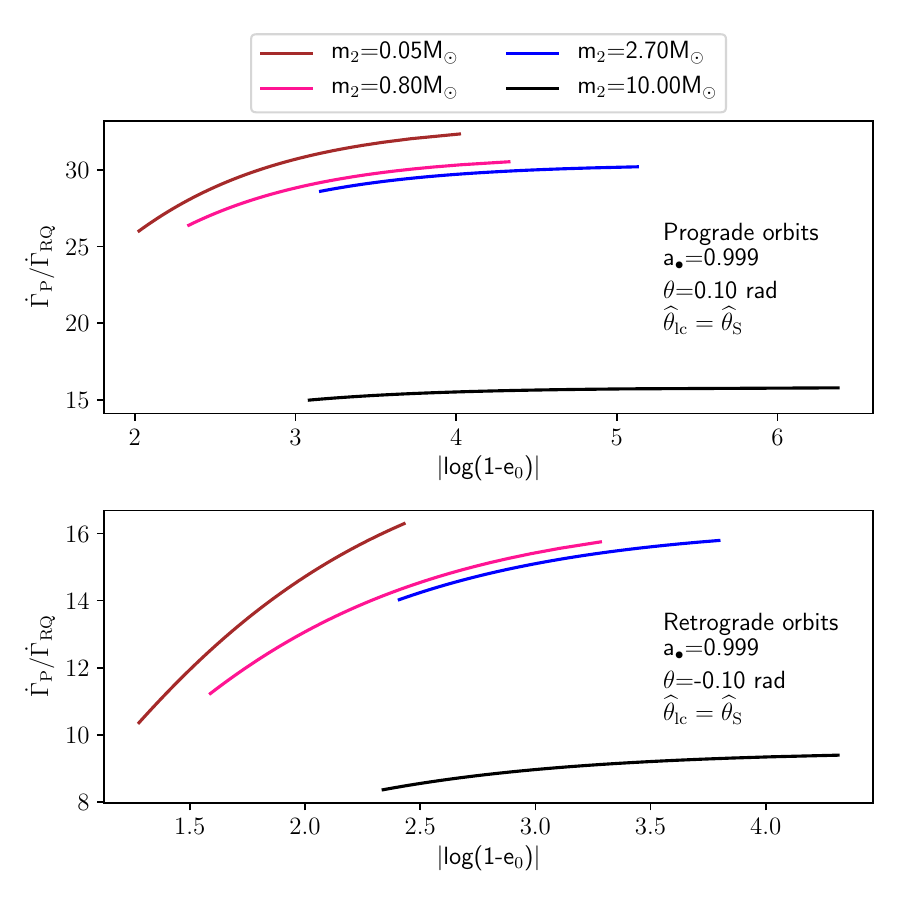}
\vspace{-15.0pt}
\caption{Ratio $\dot{\Gamma}_{\rm P}/ \dot{\Gamma}_{\rm RQ}$ as a function of the initial eccentricity for EMRIs and X-MRIs in orbits with $\theta=0.1$~rad (upper panel) and $\theta=-0.1$~rad  (lower panel) around a $4.3 \times 10^6$~M$_{\sun}$ Kerr MBH with $a_{\bullet} = 0.999$. $\dot{\Gamma}_{\rm P}$ represents the event rates obtained using Peters' formula as the merger time-scale (i.e. setting $RQ=1$ in Equation~\ref{eq:EventRate}), whereas $\dot{\Gamma}_{\rm RQ}$ are the corrected values. Both event rates are obtained with the usual loss-cone angle $\widehat{\theta}_{\rm S}$. Brown lines represent a BD ($m_2=0.05$~M$_{\sun}$) X-MRI, blue lines an inspiraling NS ($m_2=2.7$~M$_{\sun}$) EMRI, pink lines a WD ($m_2=0.8$~M$_{\sun}$) EMRI, and black lines a stellar-mass BH ($m_2=10$~M$_{\sun}$) EMRI.}
\label{fig:ratioratesRQNO}
\end{figure}

\begin{figure}
\includegraphics[width=0.5\textwidth]{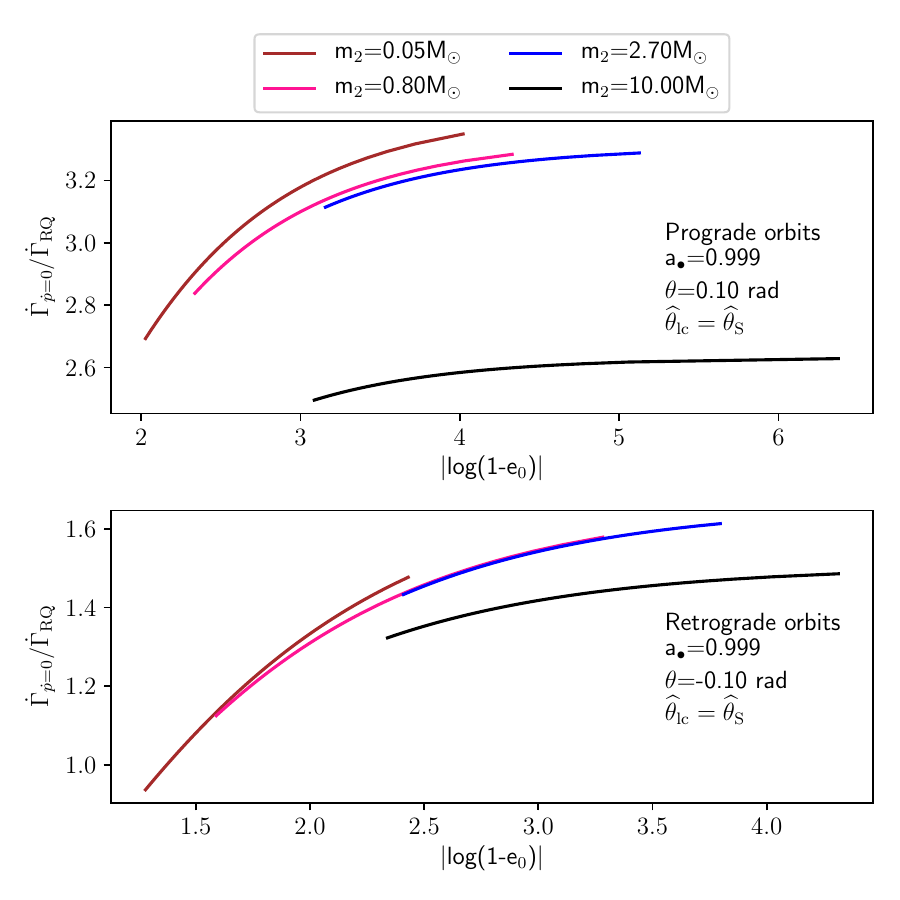}
\vspace{-15.5pt}
\caption{Same as Figure~\ref{fig:ratioratesRQNO}, but for the ratio $\dot{\Gamma}_{\dot{p}=0} / \dot{\Gamma}_{\rm RQ}$. The rates $\dot{\Gamma}_{\dot{p}=0}$ are obtained using Equation~\eqref{eq:TPapprox} as the merger time-scale.}
\label{fig:ratioratesRQ0}
\end{figure}

To investigate the effect of the correction factors in the inspiraling rates, we estimate $\dot{\Gamma}_{\rm i}$ with $\widehat{\theta}_{\rm lc}=\widehat{\theta}_{\rm S}$ in the eccentricity range given by Equations~\eqref{eq:e_max} and \eqref{eq:e_min}. We denote as $\dot{\Gamma}_{\rm P}$ the event rates without the corrections, i.e. setting $RQ=1$ in Equation~\eqref{eq:EventRate}, as $\dot{\Gamma}_{\rm RQ}$ the rates with the correction factors included, and as $\dot{\Gamma}_{\dot{p}=0}$ the event rates computed as in \citet{Pau_2013}, where Equation~\eqref{eq:TPapprox} is used as merger time-scale.

For the initial eccentricities given in Table~\ref{tab:ecc_co}, the correction factor $R$ (Equation~\ref{eq:re}) takes a value that goes from $R \sim 5$ when $e_0 \sim 0.94$, to $\sim$8 when $e\to 1$. The factor $Q$ (Equations~\ref{eq:QALSO} and ~\ref{eq:Qe}) is mainly affected by the spin and the orbital inclination; for a fixed $\theta$, the PN correction reaches its maximum value when $a_{\bullet}=0.999$. 

Figures~\ref{fig:ratioratesRQNO} and~\ref{fig:ratioratesRQ0} show the ratios $\dot{\Gamma}_{\rm P}/\dot{\Gamma}_{\rm RQ}$ and $\dot{\Gamma}_{\rm P}/\dot{\Gamma}_{\dot{p}=0}$, respectively, as a function of the initial eccentricity for the considered EMRIs and X-MRIs in prograde and retrograde orbits with $\theta = \pm 0.1$~rad, $\widehat{\theta}_{\rm lc}=\widehat{\theta}_{\rm S}$, and $a_{\bullet}=0.999$. These plots show the largest difference between the rates, as this configuration of MBH spin and orbital inclinations results in the highest (lowest) event rates in the case of prograde (retrograde) orbits. As shown in Table~\ref{tab:ratioRQ0}, the combined effect of the eccentricity evolution and the PN corrections represents an important improvement over $\dot{\Gamma}_{\rm P}$: for a central MBH with $a_{\bullet}=0.999$, $\dot{\Gamma}_{\rm P}\sim 8 - 30~ \dot{\Gamma}_{\rm RQ}$; for $a_{\bullet}=0$, $\dot{\Gamma}_{\rm P}\sim 10 - 20 ~\dot{\Gamma}_{\rm RQ}$. On the other hand, the estimate given by $\dot{\Gamma}_{\dot{p}=0}$ is between $\sim$1.3 and 2 times larger than the fully corrected value $\dot{\Gamma}_{\rm RQ}$ when $a_{\bullet}$=0, and between $\dot{\Gamma}_{\dot{p}=0} \sim 0.9 - 3 ~\dot{\Gamma}_{\rm RQ}$ for $a_{\bullet}=0.999$. 

To show the effect of the spin on $\dot{\Gamma}_{\dot{p}=0}$ and $\dot{\Gamma}_{\rm RQ}$, we compute event rates considering $a_{\bullet}=0$ and $a_{\bullet}=0.999$. For a central MBH (M$_{\rm MBH}=4.3\times10^6 $~M$_{\sun}$) with $a_{\bullet}=0$, we obtain $\dot{\Gamma}_{\rm i}^{\rm Schw}\sim 10^{-6}$--10$^{-7}$~yr$^{-1}$ for the considered EMRIs and X-MRIs. However, the event rates denoted as $\dot{\Gamma}_{\rm i}^{\rm Kerr}$ for $a_{\bullet}\neq 0$, are higher than $\dot{\Gamma}_{\rm i}^{\rm Schw}$ when the inspiraling orbits are prograde, and $\dot{\Gamma}_{\rm i}^{\rm Kerr}\lesssim \dot{\Gamma}_{\rm i}^{\rm Schw}$ if the orbits are retrograde. For $\theta=0.1$~rad, the rates $\dot{\Gamma}_{\dot{p}=0}^{\rm Kerr}$ are enhanced by a factor that can be as high as $\sim$47 with respect to $\dot{\Gamma}_{\dot{p}=0}^{\rm Schw}$. With the correction factors, we obtain that $\dot{\Gamma}_{\rm RQ}^{\rm Kerr}$ increases by a factor $\sim$23 with respect to $\dot{\Gamma}_{\rm RQ}^{\rm Schw}$ in the most extreme case ($a_{\bullet}=0.999$, $\theta=0.1$~rad). In Table~\ref{tab:DifKerrSchw}, we show the ratios $\dot{\Gamma}_{\dot{p}=0}^{\rm Kerr}/ \dot{\Gamma}_{\dot{p}=0}^{\rm Schw}$ and $\dot{\Gamma}_{\rm RQ}^{\rm Kerr}/ \dot{\Gamma}_{\rm RQ}^{\rm Schw}$.

\begin{figure}
\includegraphics[width=0.5\textwidth]{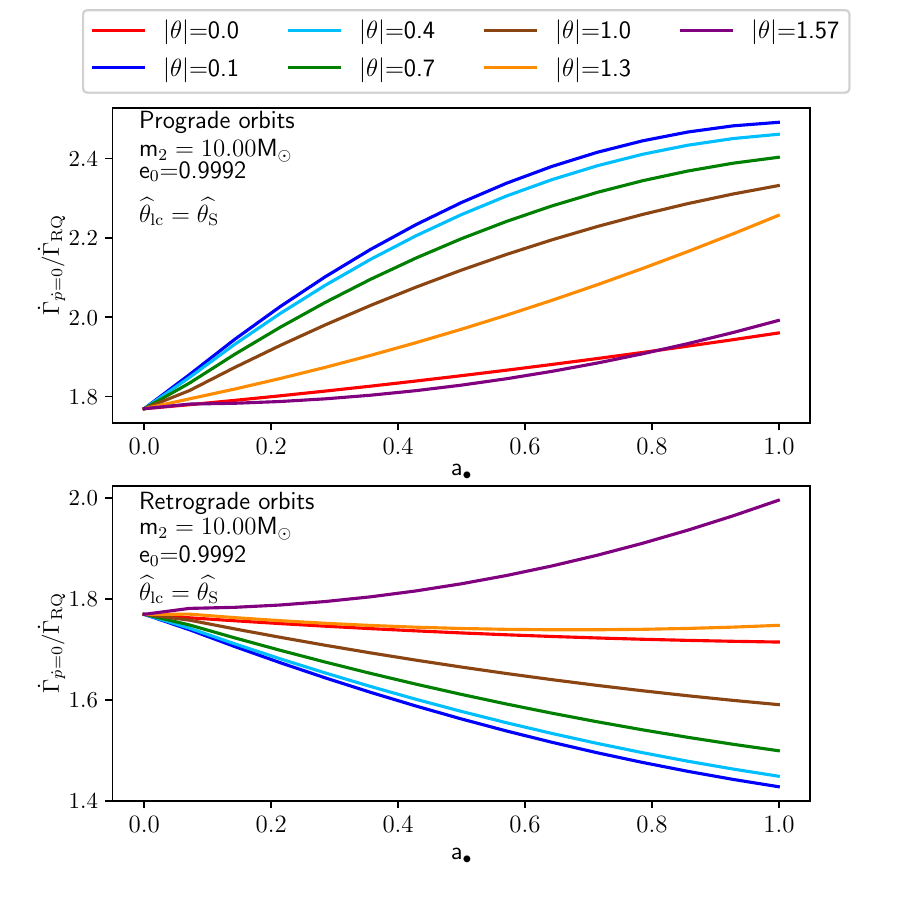}
\vspace{-15.0pt}
\caption{Ratio $ \dot{\Gamma}_{\dot{p}=0}/\dot{\Gamma}_{\rm RQ}$ as a function of the spin $a_{\bullet}$ for a BH EMRI approaching a central MBH of mass M$_{\rm MBH}= 4.3 \times 10^6$ M$_{\sun}$. The colors represent the different orbital inclinations, $\theta$, given in radians. The top (bottom) panel shows the prograde (retrograde) cases.}
\label{fig:DifRatesH}
\end{figure}

We choose the specific case of a BH EMRI approaching a central MBH (M$_{\rm MBH}=4.3\times10^6 $~M$_{\sun}$) in an orbit with $e_0=0.9992$ to show the influence of the spin and the orbital inclination. Figure~\ref{fig:DifRatesH} shows the ratio $\dot{\Gamma}_{\dot{p}=0} / \dot{\Gamma}_{\rm RQ}$ for this EMRI as a function of $a_{\bullet}$, obtained with different orbital inclinations ($\theta=[0, \pm 0.1, \pm 0.4, \pm 0.7, \pm 1.0, \pm 1.3, \pm 1.57]$~rad). In the prograde cases, the difference between $\dot{\Gamma}_{\rm RQ}$ and $\dot{\Gamma}_{\dot{p}=0}$ is larger for high spin values because the LSO shifts closer to the event horizon, and relativistic effects become more important. On the contrary, relativistic effects are weaker for retrograde orbits, as the LSO is pushed away from the event horizon. 

\begin{figure}
\centering
\includegraphics[width=0.5\textwidth]{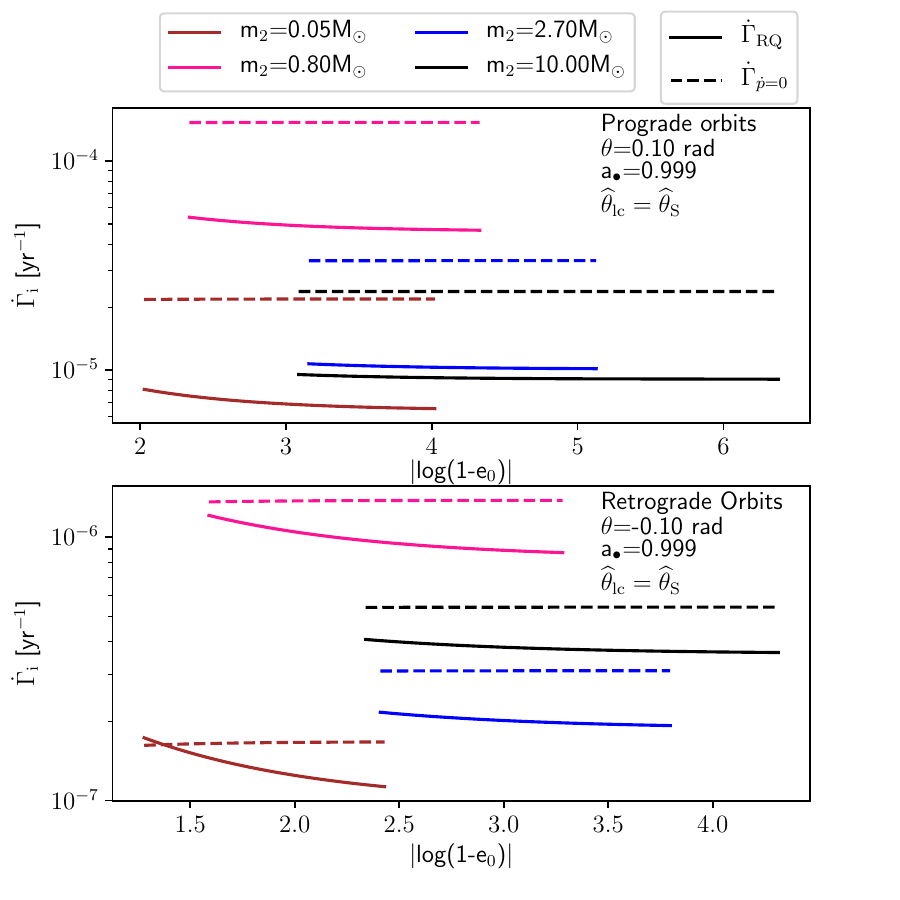}
\vspace{-15.0pt}
\caption{Inspiral event rates for prograde (upper panel) and retrograde orbits (lower panel)  with $|\theta|=0.1$~rad, around a central MBH (M$_{\rm MBH}= 4.3 \times 10^6$ M$_{\sun}$) with $a_{\bullet}=0.999$; $\dot{\Gamma}_{\dot{p}=0}$ (dotted lines) is obtained with the time-scale $T_{\dot{p}=0}$ (Equation~\ref{eq:TPapprox}), whereas $\dot{\Gamma}_{\rm RQ}$ (solid lines) comes from Equation~\eqref{eq:EventRate}. Brown lines represent a BD ($m_2=0.05$~M$_{\sun}$) X-MRI, blue lines an NS ($m_2=2.7$~M$_{\sun}$) EMRI, pink lines a WD ($m_2=0.8$~M$_{\sun}$) EMRI, and black lines a stellar-mass BH ($m_2=10$~M$_{\sun}$) EMRI.}
\label{fig:RatesH0}
\end{figure}

Finally, for the eccentricity range given in Table~\ref{tab:ecc_co}, we plot in Figure~\ref{fig:RatesH0} the rates $\dot{\Gamma}_{\rm RQ}$ and $\dot{\Gamma}_{\dot{p}=0}$ for objects with orbital inclinations $\theta = \pm 0.1$~rad approaching a Kerr MBH of mass M$_{\rm MBH}=4.3\times10^6 $~M$_{\sun}$ and  $a_{\bullet}=0.999$. Prograde WD EMRIs have the highest event rates with $\dot{\Gamma}_{\dot{p}=0}\sim 1.5\times10^{-4}$~yr$^{-1}$ and $\dot{\Gamma}_{\rm RQ}\sim 5\times10^{-5}$ yr$^{-1}$. For NS EMRIs we obtain $\dot{\Gamma}_{\dot{p}=0}\sim 3.2\times10^{-5}$ yr$^{-1}$, and $\dot{\Gamma}_{\dot{p}=0}\gtrsim 2\times10^{-5}$ yr$^{-1}$ for BH EMRIs, and BD X-MRIs. The corrected version gives $\dot{\Gamma}_{\rm RQ}\sim 1 \times10^{-5}$ yr$^{-1}$ for NS EMRIs, $\dot{\Gamma}_{\rm RQ}\gtrsim 9 \times10^{-6}$ yr$^{-1}$ for BH EMRIs, and $\dot{\Gamma}_{\rm RQ}\sim 6.5 - 8 \times10^{-6}$ yr$^{-1}$ for BD X-MRIs. 

The highest event rate of the retrograde cases is also obtained for WD EMRIs with $\dot{\Gamma}_{\dot{p}=0}\sim 1.4\times10^{-6}$ yr$^{-1}$ and $\dot{\Gamma}_{\rm RQ}\sim 1.2\times10^{-6} - 8.6 \times10^{-7}$ yr$^{-1}$. For retrograde BH EMRIs, NS EMRIs, and BD X-MRIs the event rates are $\sim 10^{-7}$ yr$^{-1}$, and $\dot{\Gamma}_{\rm RQ}\lesssim \dot{\Gamma}_{\dot{p}=0}$, with the only exception occurring at $e_0 \lesssim 0.95$, where $\dot{\Gamma}_{\dot{p}=0}\lesssim\dot{\Gamma}_{\rm RQ}$ for BD X-MRIs. These values remain approximately constant along the eccentricity range, the largest variation occurring for retrograde BD X-MRIs, where there is a difference of a factor $\sim$1.5 between the values of $\dot{\Gamma}_{\rm RQ}$ evaluated at $e_{\rm min}$ and $e_{\rm max}$.

Note that $\dot{\Gamma}_{\rm RQ}$ gives an upper limit for the event rate when $\theta=0.1$~rad, because $p_0<3 r_{\rm S}$ and the correction factor $Q$ is evaluated at $p_0=3 r_{\rm S}$, whereas for $\theta=-0.1$~rad $\dot{\Gamma}_{\rm RQ}$ contains the full PN correction as $p_0>3 r_{\rm S}$.

\begin{table}
\centering
\begin{tabular}{c|c|c} 
\hline
\addlinespace[0.04cm]
&$\dot{\Gamma}_{\rm RQ}^{\rm  Kerr}/ \dot{\Gamma}_{\rm RQ}^{\rm  Schw}$ & $\dot{\Gamma}_{\dot{p}=0}^{\rm  Kerr}/ \dot{\Gamma}_{\dot{p}=0}^{\rm  Schw}$ \T \B \\ \addlinespace[0.04cm] 
\hline
$\theta = 0.1$~rad&& \T \B \\ \hline
BD & 23.841 & 47.071 \T \B \\ \hline
WD & 23.068 & 41.817 \T \B \\ \hline
NS & 24.807 & 40.378 \T \B \\ \hline
BH & 13.371 & 19.92 \T \B \\
\hline
$\theta = - 0.1$~rad && \T \B \\ \hline
BD & 0.509 &0.350 \T \B \\ \hline
WD & 0.515 & 0.370 \T \B \\ \hline
NS & 0.500& 0.375 \T \B \\ \hline
BH & 0.573 & 0.452 \T \B \\ \hline
\end{tabular}
\caption{Ratios $\dot{\Gamma}_{\rm RQ}^{\rm  Kerr}/ \dot{\Gamma}_{\rm RQ}^{\rm  Schw}$ and $\dot{\Gamma}_{\dot{p}=0}^{\rm  Kerr}/ \dot{\Gamma}_{\dot{p}=0}^{\rm  Schw}$ obtained for WD, NS, and BH EMRIs, and a BD X-MRI in prograde and retrograde orbits, with $|\theta|= 0.1$~rad, around a central MBH with mass M$_{\rm MBH}=4.3\times10^6 $~M$_{\sun}$, and a spin value of $a_{\bullet}=0.999$ in the Kerr case.}
\label{tab:DifKerrSchw}
\end{table}

\begin{figure}
\includegraphics[width=0.5\textwidth]{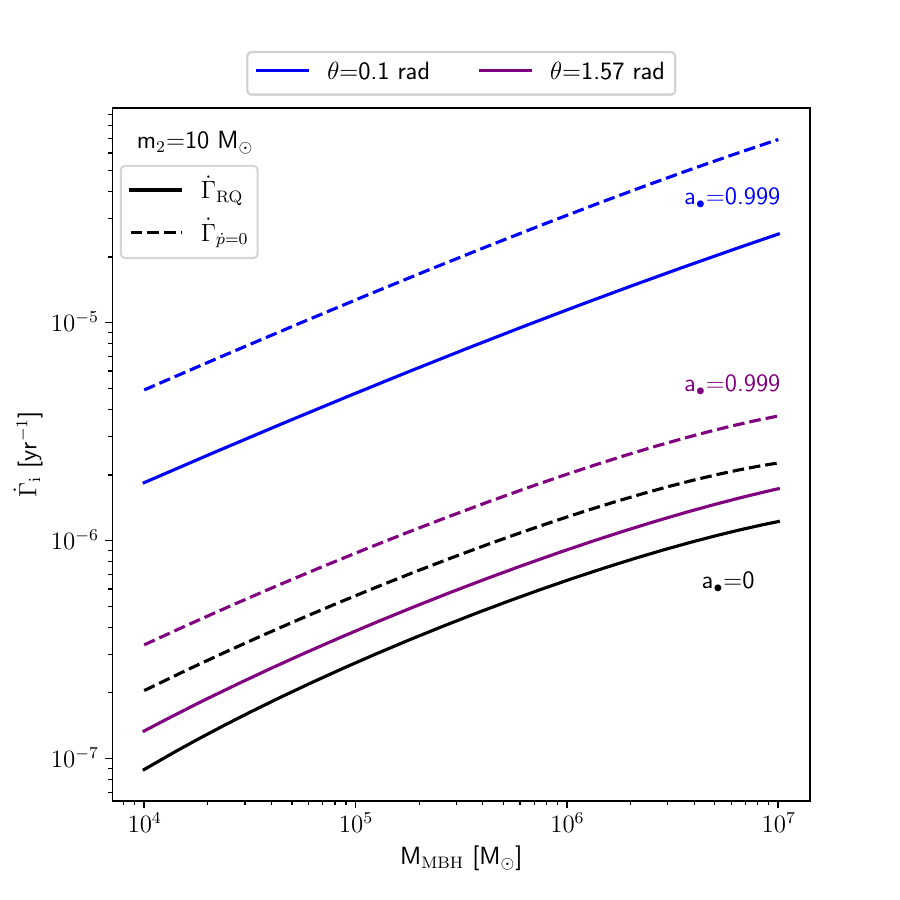}
\vspace{-15.0pt}
\caption{Event rates $\dot{\Gamma}_{\rm RQ}$, and  $\dot{\Gamma}_{\dot{p}=0}$, obtained with $\widehat{\theta}_{\rm S}$ and $e_0=e_{\rm min}$, for an EMRI composed of a BH, $m_2$=10 M$_{\sun}$, and a central MBHs with mass $M_{\rm MBH} \in [10^4, 10^7]$ M$_{\sun}$. Black lines represent the Schwarzschild case and colored lines, the Kerr case ($a_{\bullet}$=0.999). Blue lines are obtained for the orbital inclination $\theta=0.1$~rad, purple lines, for $\theta=1.57$~rad. }
\label{fig:ratesMBHs}
\end{figure}

\begin{figure}
\includegraphics[width=0.5\textwidth]{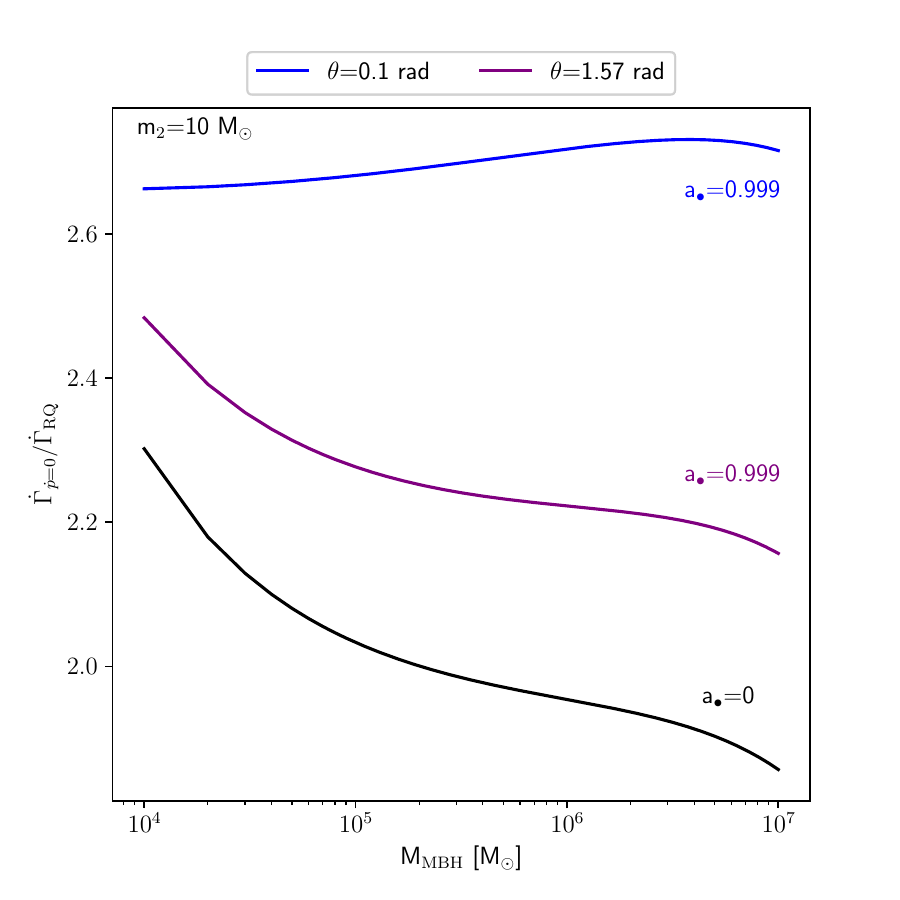}
\vspace{-15.0pt}
\caption{Ratio $\dot{\Gamma}_{\dot{p}=0}/\dot{\Gamma}_{\rm RQ}$, for a BH EMRI with a central MBH with mass $M_{\rm MBH} \in [10^4, 10^7]$ M$_{\sun}$. Black lines represent the Schwarzschild case and colored lines, the Kerr case ($a_{\bullet}$=0.999). Blue lines are obtained for the orbital inclination $\theta=0.1$~rad, purple lines, for $\theta=1.57$~rad. The event rates are obtained considering $\widehat{\theta}_{\rm S}$ and $e_0=e_{\rm min}$.}
\label{fig:ratesratioMBHs}
\end{figure}

LISA will be able to detect EMRIs and X-MRIS if the mass of the central MBH is between $\sim 10^4$~M$_{\sun}$ and $\sim 10^7$~M$_{\sun}$. For lower MBH masses, the GWs amplitude would be very low and the source would need to be located within a few Gpc to be detected \citep{2004CQG, Pau_2007}. On the other hand, if the mass of the MBH is higher than $\sim 10^7$~M$_{\sun}$, the signal's frequency would be too low to be detected. \\ 
Black holes at the low-mass end ($\sim 10^{4-5}$~M$_{\sun}$)  can be identified as intermediate-mass black holes (IMBH). Although their existence and the validity of the scaling relations remain uncertain, kinematic observations of globular clusters (GC) and dwarf galaxies seem to indicate the presence of IMBHs in the central region of these stellar systems \citep{Lutzgendorf_2013, Lutzgendorf_2014, Tremou_2018, Reines_2015, 2020_dwarfgal}. \\
The description presented in this work can be extended to systems with $M_{\rm MBH}\in (10^{4}$~M$_{\sun},10^{7}$~M$_{\sun})$ assuming that scaling relations between the central MBH and the host stellar system holds for this MBH mass range. In that case, the critical semimajor axis scales as
\begin{align}
a_{\rm crit}& \sim  5.33 \times 10^{-1}\, \text{pc} \, \times \mathcal{R}\, \mathcal{Z} \left(\mathcal{N}\,\hat{e}\right)^{-4/5}\times \label{eq:Scaleacrit} \\
&\times\left(\frac{m_{\rm BH}}{10M_{\sun}}\right)^{-8/5}  \left(\frac{10M_{\sun}}{m_{2}}\right)^{-4/5}  \left[ \frac{4.3\times 10^6 M_{\sun}}{M_{\rm MBH}}\right]^{-4/5}, \nonumber
\end{align}
where we take an eccentricity of $0.9992$ so that 
\begin{align}
\hat{e}\approx  \left( \frac{1}{1.28}\right) \frac{f(e_0)}{(1-e_0^2)(1-e_0)^{5/2}},   \nonumber
\end{align}
and
\begin{align}
\mathcal{R}&= \left(\frac{R_{\rm h}}{1 \text{pc}}\right), \, \mathcal{N}=\frac{N_0}{12000} \frac{\text{ln}(\Lambda)}{13},\nonumber \\
\mathcal{Z}&= (R\,Q)^{-4/5} \mathcal{W}(\theta, a_{\bullet})^{-2}. 
\end{align}
The event rate for $m_2=m_{\rm BH}=10$~M$_{\sun}$, $\beta=\gamma=7/4$, and $\widehat{\theta}_{\rm lc}=\widehat{\theta}_{\rm S}$, scales as
\begin{align}
\dot{\Gamma}_{\rm i}\sim  &7.76\times 10^{-8}\, \text{yr}^{-1} \, \times \mathcal{R}^{-4/5} \mathcal{N}^{1/5} \times  \label{eq:ScaleRate}  \\
&\times \left(\frac{M_{\rm MBH}}{4.3\times 10^6 M_{\sun}}\right)^{3/5} \Bigg\lbrace \mathcal{Z} \, \hat{e}^{-4/5}  \times  \Bigg [ 96 \,+ \nonumber \\
&+ \text{ln} \left( \mathcal{Z} \, \mathcal{R} \left(\mathcal{N} \hat{e} \right)^{-4/5} \left( \frac{M_{\rm MBH}}{4.3\times 10^6 M_{\sun}} \right)^{-1/5} \right) \Bigg ] \nonumber \\
&- 9.6\times 10^{-6}  \mathcal{N}^{-4/5} \left(\frac{M_{\rm MBH}}{4.3\times 10^6 M_{\sun}}\right)^{8/5}\times \nonumber \\
&\times \left[ 89+\text{ln}\left( \mathcal{R} \left(\frac{M_{\rm MBH}}{4.3\times 10^6 M_{\sun}}\right)^{-8/5}\right)\right]\Bigg\rbrace. \nonumber 
\end{align}
Note that to obtain the correct value for $a_{\rm crit}$ and $\dot{\Gamma}_{\rm i}$, the term $\mathcal{Z}$ has to be evaluated. Neither the function $\mathcal{W}(\theta, a_{\bullet})$ nor the correction factors $RQ$ (Equations~\ref{eq:re},~\ref{eq:QALSO},~\ref{eq:Qe}) depend on the mass of the MBH or $m_2$. Therefore, the effect of the term $\mathcal{Z}$ in the last equation is fixed for a given set of $a_{\bullet}$, $\theta$, and $e_0$; it decreases the event rates by a factor that remains between $\lesssim 0.9$ and $\gtrsim 3$ for the different orbital inclinations and spin value.\\
Figure \ref{fig:ratesMBHs} shows $\dot{\Gamma}_{\rm RQ}$, and $\dot{\Gamma}_{\dot{p}=0}$, obtained for a BH EMRI with orbital inclinations $\theta=1.57$ and $\theta=0.1$~rad, around a central MBH of mass $M_{\rm MBH} \in [10^4, 10^7]$ M$_{\sun}$, $a_{\bullet}$=[0, 0.999], and $e_0=e_{\rm min}$ (Equation~\ref{eq:e_min}). To compute the value of $e_0$, we assume that the $M-\sigma$ relation for the velocity dispersion described in  \citet{Tremaine2002}, $\sigma_0 \sim 200~(M_{\rm MBH}/10^8 M_{\sun})^{1/4}$ km/s, holds for the considered $M_{\rm MBH}$ masses. The event rate decreases with the mass of the central MBH; also, as the value of $e_{\rm min}$ depends on $M_{\rm MBH}$ and $R_{\rm h}$, EMRIs in the low-mass are more eccentric compared to EMRIs formed around the most massive central black holes; a similar behavior occurs if we take $e_0=e_{\rm max}$ since the magnitude of $a_{\rm crit}$ also decreases with the mass of the central MBH.   \\
Figure \ref{fig:ratesratioMBHs} shows the ratio $\dot{\Gamma}_{\dot{p}=0}/\dot{\Gamma}_{\rm RQ}$ for the same EMRI configuration. As the effect of the spin is weak on highly inclined orbits ( $\theta=1.57$~rad), the effect of the eccentricity becomes more important, increasing the ratio $\dot{\Gamma}_{\dot{p}=0}/\dot{\Gamma}_{\rm RQ}$ as $M_{\rm MBH}\to 1\times 10^4$. For $M_{\rm MBH}=1\times 10^7$M$_{\sun}$, the eccentricity value is $\sim $0.995, so the effect of the $RQ$ corrections is weaker than for the less massive MBHs where $e_0\to0.999999$. On the other hand, for $\theta=0.1$~rad the influence of the spin dominates over the eccentricity evolution and PN effects are stronger; $\dot{\Gamma}_{\dot{p}=0}/\dot{\Gamma}_{\rm RQ}\sim 2.6 - 2.7$ along the MBH's mass range.

\subsection{The effect of the loss-cone angle}  

The shift in the LSO position can reduce or increase the magnitude of the loss-cone angle, modifying the phase-space volume that places the pericentre of an orbit inside the LSO. As $a_{\bullet}\to 1$, the Kerr loss-cone angles $\widehat{\theta}_{\mathcal{W}}$ and $\widehat{\theta}_{\rm K}$ (Equations~\ref{eq:LossConeW} and \ref{eq:LossConeK}) deviate more from the Schwarzschild loss-cone angle $\widehat{\theta}_{\rm S}$. However, the change in the event rates is small even for high spin values, as $\dot{\Gamma}_{\rm i} \propto \ln (\widehat{\theta}^{-2}_{\rm lc})$.

The factor $Q$ indicates that the influence of $a_{\bullet}$ and $\theta$ is not as large as the one obtained when the function $\mathcal{W}(\theta,a_{\bullet})$ is implemented (see Table~\ref{tab:DifKerrSchw}), so it is no surprise that the rates $\dot{\Gamma}_{\dot{p}=0}$ are more affected by $\widehat{\theta}_{\mathcal{W}}$ and $\widehat{\theta}_{\rm K}$ compared to $\dot{\Gamma}_{\rm RQ}$. 

\begin{figure}
\includegraphics[width=0.5\textwidth]{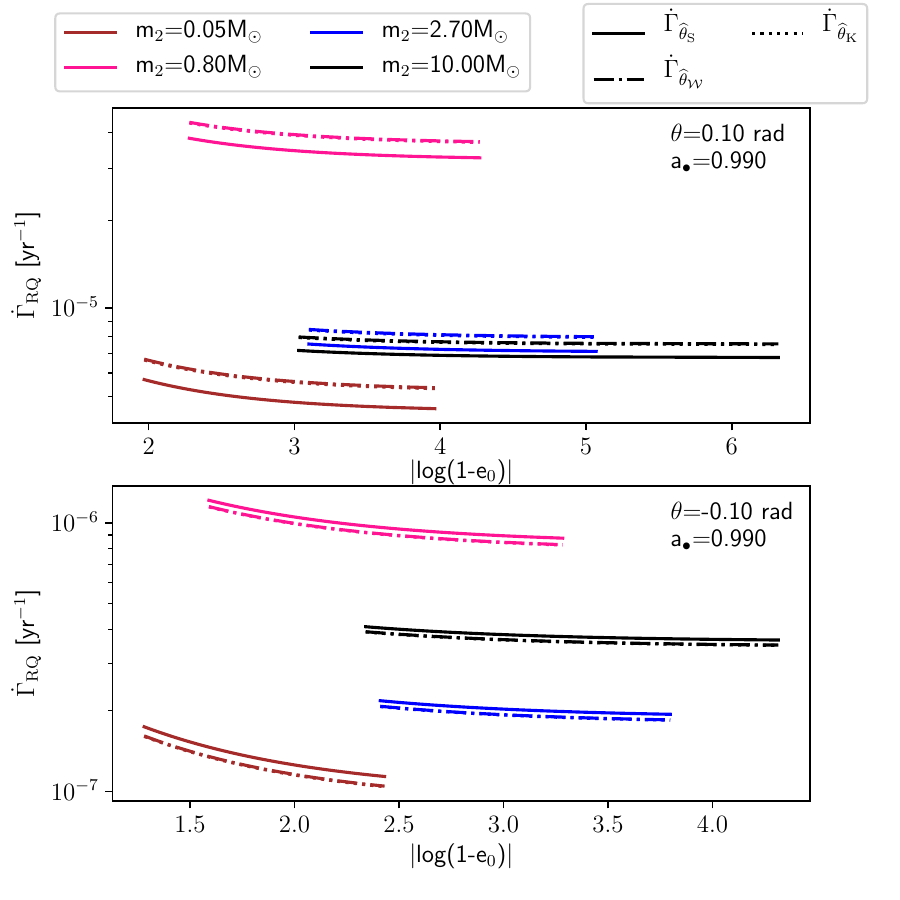}
\vspace{-15.0pt}
\caption{Event rates $\dot{\Gamma}_{\rm RQ}$ computed with the different loss-cone angles, as a function of $e_0$ for EMRIs and X-MRIs with $|\theta|=0.1$~rad in prograde (upper panel), and retrograde orbits (lower panel), around a MBH of mass M$_{\rm MBH}= 4.3 \times 10^6$~M$_{\sun}$, and $a_{\bullet}=0.99$. The subscript indicates which loss-cone angle is used to obtain the event rates; $\widehat{\theta}_{\rm S}$ is given by Equation~\eqref{eq:LossConeS}, $\widehat{\theta}_{\mathcal{W}}$ is the Kerr loss-cone angle given by Equation~\eqref{eq:LossConeW}, and $\widehat{\theta}_{\rm K}$ is given by Equation~\eqref{eq:LossConeK}.}
\label{fig:DifLCrates}
\end{figure}

By computing $\dot{\Gamma}_{\rm RQ}$ and $\dot{\Gamma}_{\dot{p}=0}$ with $\widehat{\theta}_{\rm lc}=\widehat{\theta}_{\rm S}$, $\widehat{\theta}_{\mathcal{W}}$, and $\widehat{\theta}_{\rm K}$ for a spin value of $a_{\bullet}=0.99$ that guarantees the accuracy within 5 per cent of $J_{\rm crit}$ (Equation~\ref{eq:LcritkKerr}), we find that the asymmetric effect of the MBH spin is still noticeable. It can be seen in  Figure~\ref{fig:DifLCrates}, where we show the corrected event rates computed with the different loss-cone angles -- $\dot{\Gamma}_{\widehat{\theta}_{\rm S}}$, $\dot{\Gamma}_{\widehat{\theta}_{\mathcal{W}}}$, and $\dot{\Gamma}_{\widehat{\theta}_{\rm K}}$ -- as a function of $e_0$ for the considered EMRIs and X-MRIs. 

In Section~\ref{sec:LossConeAngle}, we showed that, for prograde orbits around a Kerr MBH, $\widehat{\theta}_{\mathcal{W}}\sim \widehat{\theta}_{\rm K}$, and that, for high spin values, $\widehat{\theta}_{\mathcal{W}}\lesssim$ 0.5 $\widehat{\theta}_{\rm S}$. This reduction in the loss-cone angle increases the EMRI and X-MRI event rates $\dot{\Gamma}_{\dot{p}=0}$ by a factor $\sim$1.2 compared to the Schwarzschild case. For objects in retrograde orbits, $\dot{\Gamma}_{\dot{p}=0}$ are reduced by a factor $\sim$0.8 due to the small increase in the magnitude of the Kerr loss-cone angle. In the case of $\dot{\Gamma}_{\rm RQ}$, the Kerr loss-cone angle changes the rates estimate by a factor $\sim$0.9 to $\sim$1.1, which is still negligible. In Table~\ref{tab:ratioLC}, we give the ratio between these rates.

\begin{table}
\centering
\begin{tabular}{c||c|c||c|c}
\hline
 \addlinespace[0.05cm]
$\dot{\Gamma}_{\rm RQ}$&  \multicolumn{2}{c||}{$\dot{\Gamma}_{\widehat{\theta}_{\mathcal{W}}}/\dot{\Gamma}_{\widehat{\theta}_{\rm S}}$}  &\multicolumn{2}{c}{$\dot{\Gamma}_{\widehat{\theta}_{\rm K}}/\dot{\Gamma}_{\widehat{\theta}_{\rm S}}$} \T \B \\ \addlinespace[0.05cm] \hline
 $\theta$ [rad] &  0.1$ $  &  $-0.1$ & 0.1  & $-0.1$ \T \B \\ \hline\hline
\addlinespace[0.15cm]
BD & 1.17 & 0.92 & 1.16 & 0.92 \T \B \\ \hline
WD & 1.13 & 0.94& 1.12 & 0.94 \T \B \\ \hline
NS & 1.12 & 0.95& 1.11 & 0.95 \T \B \\ \hline
BH & 1.11 & 0.95 & 1.10 & 0.95 \T \B \\ \hline
\addlinespace[0.5cm]
\hline
\addlinespace[0.05cm]
$\dot{\Gamma}_{\dot{p}=0}$&  \multicolumn{2}{c||}{$\dot{\Gamma}_{\widehat{\theta}_{\mathcal{W}}}/\dot{\Gamma}_{\widehat{\theta}_{\rm S}}$} &  \multicolumn{2}{c}{$\dot{\Gamma}_{\widehat{\theta}_{\rm K}}/\dot{\Gamma}_{\widehat{\theta}_{\rm S}}$} \T \B \\ \addlinespace[0.05cm] \hline
 $\theta$ [rad] &  0.1$ $  &  $-0.1$ & 0.1  & $-0.1$ \T \B \\ \hline\hline
\addlinespace[0.15cm]
\hline
BD & 1.22& 0.89 & 1.21 & 0.88 \T \B \\ \hline
WD & 1.16 & 0.93&1.15 & 0.92 \T \B \\ \hline
NS & 1.14 & 0.94& 1.13 & 0.93 \T \B \\ \hline
BH & 1.13 & 0.94& 1.12 & 0.94 \T \B \\ \hline
\end{tabular} 
\caption{Comparison of event rates computed with $\widehat{\theta}_{\rm S}$ and the Kerr loss-cone angles $\widehat{\theta}_{\mathcal{W}}$ and $\widehat{\theta}_{\rm K}$ for each inspiraling object. We take $\theta=\pm 0.1$~rad, M$_{\rm MBH}=4.3 \times 10^6$~M$_{\sun}$, and $a_{\bullet}=0.99$. The upper section shows the change induced by the Kerr loss-cone angles in $\dot{\Gamma}_{\rm RQ}$ and the lower part shows the change in $\dot{\Gamma}_{\dot{p}=0}$. }
\label{tab:ratioLC}
\end{table}


\section{DISCUSSION AND CONCLUSIONS}\label{sec:conclussions}

The spin of the MBH and the orbital inclination of the inspiraling object can not be ignored; these quantities can affect the event rates in two forms. Firstly, through the pericentre of an inspiraling orbit: as we fix the pericentre at $r_{\rm LSO}$, a shift in the LSO position changes the value of the critical semimajor axis and the integration volume of Equation~\eqref{eq:Rates}, significantly enhancing the event rates of prograde orbits, and slightly reducing the event rates in the retrograde cases, as the effect of the MBH spin is not symmetric. Secondly, through the loss-cone: its value also depends on $r_{\rm LSO}$ and determines the set of velocity vectors that take an object to a direct plunge. We give two expressions, $\widehat{\theta}_{\mathcal{W}}$ and $\widehat{\theta}_{\rm K}$, to obtain a loss-cone angle that accounts for spin effects. Both versions of the Kerr loss-cone angle give similar results and, as $a_{\bullet} \to$ 1, $\widehat{\theta}_{\mathcal{W}}$ and $\widehat{\theta}_{\rm K}$ deviate more from the Schwarzschild case $\widehat{\theta}_{\rm S}$. However, we find that the influence of the MBH spin added through the pericentre condition, $p_0 = r_{\rm LSO}$, already contains the most relevant effects regarding the event rates, so that implementing $\widehat{\theta}_{\mathcal{W}}$ or $\widehat{\theta}_{\rm K}$ changes the event rates by a factor that ranges between 0.9 and 1.2, which does not produce a significant impact on the rate estimates.

We obtained event rates for EMRIs and X-MRIs by implementing three different merger time-scales, $T_{\rm P}$, $T_{\rm RQ}$, and $T_{\dot{p}=0}$. Peters' formula, $T_{\rm P}$, overestimates the energy loss by GWs and fails to give an accurate merger time-scale; this can be avoided by including eccentricity evolution and post-Newtonian corrections through the correction factors $R$ and $Q$; the resulting time-scale, $T_{\rm RQ}$, is longer than $T_{\rm P}$ and produces the best merger time-scale estimate for arbitrary eccentricities, orbital inclinations, and MBH spin values. The alternative formulation $T_{\dot{p}=0}$ gives a reliable estimate of the merger time-scale, $T_{\dot{p}=0} \lesssim T_{\rm RQ}$, in the context of EMRIs and X-MRIs. However, for arbitrary values of $e_0$, $\theta$, or $a_{\bullet}$, its accuracy can not be guaranteed.

We have shown that for the eccentricity range and pericentre distances expected for EMRIs and X-MRIs ($e_0> 0.9$, $p_0= r_{\rm LSO}$), implementing $T_{\rm GW}=T_{\rm P}$ results in unreliable event rates estimates. $\dot{\Gamma}_{\rm P}$ are artificially enhanced by a factor that ranges between $\sim$8 to 30 compared to the corrected values $\dot{\Gamma}_{\rm RQ}$. On the other hand, the estimates given by $\dot{\Gamma}_{\dot{p}=0}$, which include the influence of the MBH spin and the orbital inclination through the function $\mathcal{W}(\theta, a_{\bullet})$, differ from $\dot{\Gamma}_{\rm RQ}$ by a factor between 0.9 and 3.

We conclude that both the Kerr loss-cone and the $RQ$ corrections to Peters' time-scale do not have a dramatic impact on the event rates for EMRIs or X-MRIs, when compared to the high-eccentricity approach of \citet{Pau_2013}. However, this work considers only the dynamical and relativistic aspects of the EMRIs and X-MRIs formation and considers the galactic nucleus of the Milky Way as a representative example of the galaxies that could harbour potential inspiral sources. If EMRIs and X-MRIs in Nature happened to form at very low eccentricities, or if environmental effects (not included in this work; e.g. torques induced by background gas) can induce a significant reduction of the initial eccentricity of EMRIs and X-MRIs ($e_0\lesssim 0.8$), it would be necessary to implement the eccentricity-evolution and PN corrections to obtain accurate event rate estimates. In any case, all event rates for MBH binaries and stellar-mass binaries should be revisited using our improvements because their eccentricities will span through all possible values. Furthermore, our description holds for MBH with masses between $\sim 10^{4}$M$_{\sun}$ and $\sim 10^{7}$M$_{\sun}$, covering the mass interval detectable by LISA.

\section*{ACKNOWLEDGMENTS}
VVA acknowledges support from CAS-TWAS President's PhD Fellowship Programme of the Chinese Academy
of Sciences \& The World Academy of Sciences. PAS acknowledges support from the Ram{\'o}n y Cajal Programme of the Ministry
of Economy, Industry and Competitiveness of Spain, as well as the financial
support of Programa Estatal de Generación de Conocimiento (ref.
PGC2018-096663-B-C43) (MCIU/FEDER). This
work was supported by the National Key R\&D Program of China (2016YFA0400702)
and the National Science Foundation of China (11721303, 11873022 and 11991053).
PRC, LM, and LZ acknowledge support from the Swiss National Science Foundation under the Grant 200020\_178949. EB acknowledges support from the European Research Council (ERC) under the European Union's Horizon 2020 research and innovation program ERC-2018-COG under grant agreement N.~818691 (B~Massive). 

\section*{Data Availability Statement}
The data underlying this article will be shared on reasonable request to the corresponding author.

\bigskip

\small
\bibliographystyle{mnras}
\bibliography{RevisedRates.bib}
\normalsize 


\bsp 
\label{lastpage}
\end{document}